\documentclass{article}

\usepackage{arxiv}

\usepackage[utf8]{inputenc} % allow utf-8 input
\usepackage[T1]{fontenc}    % use 8-bit T1 fonts
\usepackage{booktabs}       % professional-quality tables
\usepackage{nicefrac}       % compact symbols for 1/2, etc.
\usepackage{microtype}      % microtypography
\usepackage{doi}

\usepackage[english]{babel}
\usepackage{amsmath}
\usepackage{amstext}
\usepackage{graphicx}%
\usepackage{amsfonts}%
\usepackage{amssymb}
\usepackage{dsfont}
\usepackage{rotating}
\usepackage{xy}
\usepackage{pdflscape}
\usepackage{url}
\usepackage[round,authoryear]{natbib}
\usepackage{mathrsfs}
\usepackage{setspace}
\usepackage{caption}
\usepackage{multirow}

\newcommand{\bs}{\mathbf{s}}

\newcommand{\bX}{\mathbf{X}}

\newcommand{\bZ}{\mathbf{Z}}
\newcommand{\bz}{\mathbf{z}}

\newcommand{\bbeta}{ {\boldsymbol \beta} }
\newcommand{\bdelta}{ {\boldsymbol \delta} }

\newcommand{\bSigma}{\boldsymbol{\Sigma}}

\date{January 11, 2022}	% Here you can change the date presented in the paper title
\title{Modeling First Arrival of Migratory Birds using a Hierarchical Max-infinitely Divisible Process}
\author{Dhanushi A. Wijeyakulasuriya \\
	Microsoft Corporation \\
	1 Microsoft Way, Redmond, WA, 98052 \\
	\And
	Ephraim M. Hanks \\
	Department of Statistics \\
	Pennsylvania State University\\
	University Park, PA, 16802 \\
	\AND
	Benjamin A. Shaby \\
	Department of Statistics\\
	Colorado State University\\
	Fort Collins, CO, 80523
}

\renewcommand{\shorttitle}{Modeling First Arrivals with Max-ID Processes}

\hypersetup{
pdftitle={Modeling First Arrivals with Max-ID Processes},
pdfauthor={Dhanushi A. Wijeyakulasuriya, Ephraim M. Hanks, Benjamin A. Shaby},
pdfkeywords={Extreme value theory; Max-stable process; Spatial extremes},
}

\begin{document}
\maketitle

\begin{abstract}
	Humans have recorded the arrival dates of migratory birds for millennia, searching for trends and patterns.  As the first arrival among individuals in a species is the realized tail of the probability distribution of arrivals, the appropriate statistical framework with which to analyze such events is extreme value theory.  Here, for the first time, we apply formal extreme value techniques to the dynamics of bird migrations.  We study the annual first arrivals of Magnolia Warblers using modern tools from the statistical field of extreme value analysis. Using observations from the eBird database, we model the spatial distribution of Magnolia Warbler arrivals as a max-infinitely divisible process, which allows us to spatially interpolate observed annual arrivals in a probabilistically-coherent way, and to project arrival dynamics into the future by conditioning on climatic variables.
\end{abstract}

% keywords can be removed

\section{Introduction}
Patterns in spring bird migrations are key indicators of ecosystem responses to climate pressures. For millenia, observers have used the annual first arrival of a given species to demarcate the migration \citep{Lincoln1935}. The date of first arrival is a statistical extreme value; that is, the first individual from among a large population to reach a given location is exactly the realized tail of the probability distribution of all arrivals.  The statistical theory of extreme values \citep{beirlant2004statistics, Davison2019} is thus ideally suited for modelling spring first arrivals, but this theory has never been used for modeling first arrivals. Ours is the first such study to deploy modern techniques from the statistics of extremes to model this phenomenon. In this analysis, we use state-of-the-art tools for spatial extremes to model and predict the first arrival of migratory birds. We show that using these methods allows for principled inference on the relationship between landscape or climatic variables and first arrival times, predictions at unobserved locations during past years, and predictions of future first arrivals under climate model projections.

We frame the problem of modeling the first arrival of migratory birds as a problem of modeling spatial extremes. We apply this method to study and predict the Spring arrival of Magnolia Warblers (\textit{Setophaga  magnolia}) in the Northeast portion of the United States from 2004--2019 using data from the eBird database \citep{Ebirddata}, a citizen science website. By using hierarchical spatial extreme value models, we are able to obtain conditional predictions of first arrival dates at locations without eBird observations. 

Previous studies have investigated the mismatch between the arrival of migratory birds and other aspects of the onset of Spring \citep{kolecek_shifts_2020, kullberg_change_2015, tottrup_local_2010, Jonzen2006}. Most used linear regression or similar methods that regress a measure of first arrival time on covariates related to the onset of Spring \citep{Moller16195, mayor_increasing_2017, Gunnarsson2011}. Some methods incorporated multiple species using a random effect \citep{kolecek_shifts_2020, kullberg_change_2015}. \cite{ambrosini2014} used a binomial conditional auto-regressive mixed model, and is the only instance of which we are aware that employed a formal spatial model for first arrivals. There is little uniformity in the definition of first arrival times in existing literature. \cite{palm2009}, \cite{Gunnarsson2011}, and  \cite{kolecek_shifts_2020} calculated first arrival date for a given year to be the mean of first arrival dates recorded by observers averaging over all geographical locations. Another approach is to fit logistic or cumulative log log functions to estimate the first arrival time as the inflection point or specific percentiles \citep{mayor_increasing_2017, ambrosini2014}. \cite{zaifman2017} used a set of heuristics to filter out noisy data points from the eBird database to identify first arrival times.  All of these approaches modeled first arrival times using standard, mean-focused statistical models, rather than modeling first arrival times as the extreme values that they are.

We construct our extreme value models by conceptualizing the geographical map of first arrivals as a spatial field block minima.  That is, at each point in space, a subset of the population of Magnolia Warblers visits, with each member of that population arriving at a particular time.  In the parlance of extreme value statistics, this collection of arrival times is referred to as a ``block'', and the earliest arrival time is the block minimum.  By singling out the first arrival time at each of many spatial locations, we arrive at a spatial field of block minima.  This is exactly the structure of data that is the subject of the statistical study of spatial extremes.  The enterprise of modeling spatial extremes has most often focused on extreme weather events like extreme precipitation, temperature, and wind \citep{reich2012,shaby2012,huser2019,reich2019}.  % For recent reviews in this field, see \citet{Davison2019} and \citet{huser2021}. 
Extreme value methods have only been rarely used in ecological studies \citep{wijeyakulasuriya2019}.

Widely-used models for spatial statistics based on Gaussian processes are appropriate for modeling spatial events in the bulk of a distribution, but they do not do well for spatial extremes because they have rapidly-vanishing tail dependence which is usually not realistic for extremes \citep{bopp2020}.  Furthermore, they do not posses the max-infinite divisibility (max-id) property, which we will argue is necessary for any coherent model of block minima.  Instead of using Gaussian processes, we model the timing of the Magnolia Warbler Spring migration using the hierarchical model of \citet{bopp2020}, which is max-id (and therefore appropriate for block minima like first arrival times), has flexible tail dependence properties that can be learned from data, and decomposes in such a way as to make computing tractable on large datasets.

\begin{figure}
	\begin{center}
		%\vspace*{-4cm}
		\includegraphics[width=1.0\linewidth,clip=true,trim={0 290 0 25}]{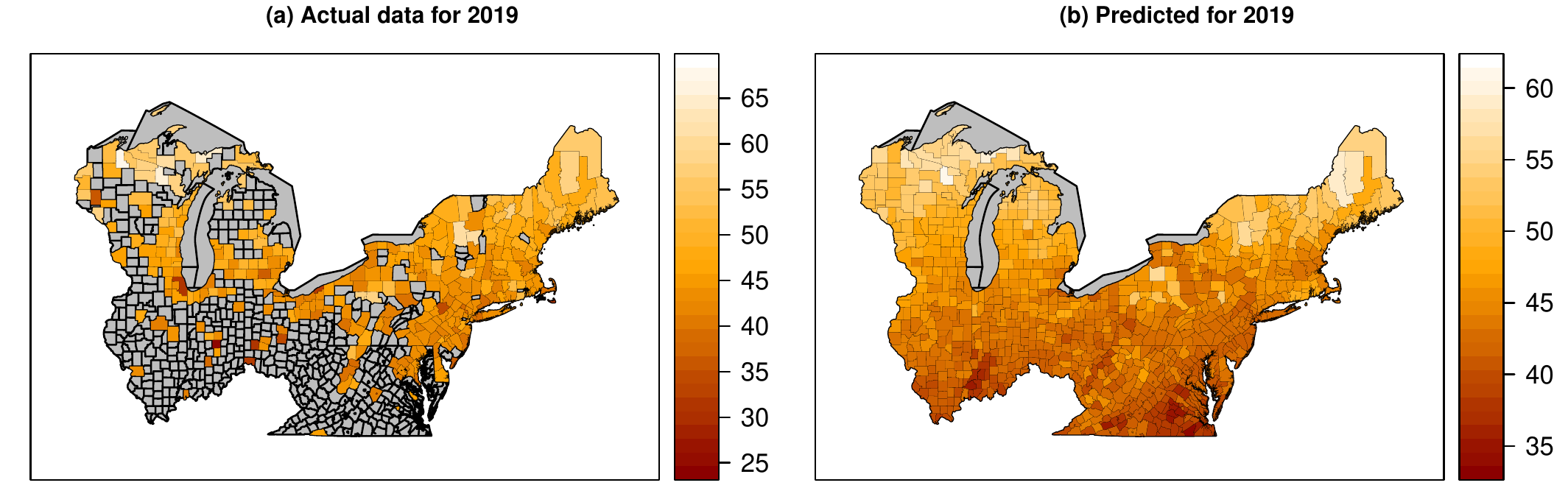} 
		\captionof{figure}{(a) First arrival data for 2019. Missing values are shown in grey. (b) Predicted First arrival for 2019.}\label{fig:2019_preds} 
	\end{center}
\end{figure}

The hierarchical modeling approach also makes it easy to incorporate covariates.  We fit fifteen years of first arrival data from eBird, regressing model parameters on topographic, landcover, demographic, and climatalogical predictors. We then create predictive maps of first spring arrival for the Magnolia Warbler in the northeast United States (Section 3). For an example of eBird data and the model's predicted map, see Figures \ref{fig:2019_preds}(a) and \ref{fig:2019_preds}(b). We then use climate model data from the fifth phase of the Coupled Model Intercomparison Project (CMIP5) for years 2151--2200 to demonstrate how these models can be used to make future predictive maps of first arrival (See Figures  \ref{fig:meanandsdclimate}(1) and \ref{fig:meanandsdclimate}(b)).

\subsection{Data Processing}

We downloaded bird sightings data for the Magnolia Warbler from the citizen data website eBird \citep{Ebirddata}. We focused our study on the northeast US and only considered years from 2004--2019 due to the relative lack of data before that period. Since we were looking at Spring arrivals, we included sightings from March 20 to July 20 of each year. 
%
%We calculated arrival time at the county level (869 counties in total) as the number of days since March 20, and the first arrival time as the minimum of these arrival times. In order to minimize the influence of potentially aberrant observations, we calculated first arrival time only in counties that had at least 12 sightings in a given year. There is a large proportion of missing data (80\%) in the overall dataset. See Figure \ref{fig:2019_preds} (a) for the observed first arrival times for 2019. Counties colored in grey do not have first arrival observations in 2019.

We included spatial and climate covariates to model variation in first arrival times. The spatial covariates are longitude, latitude, elevation \citep{elevatr_R}, forest cover \citep{forestcov}, proportion of water, and population density \citep{census2010}. Longitude and latitude covariates were taken to be the centroid of each county. All other covariates are the spatial average over the county. The climate covariates are temperature anomaly \citep{tempanom} and North Atlantic Oscillation (NAO) \citep{NAO} for the month of March of each year.  For more details about the first arrival data and spatial and climate covariates, see %Supplementary Information (SI)
Appendix \ref{appendix:data-processing-all}.

\section{Hierarchical Max-Infinitely Divisible Spatial Process Model}

Models for extreme values are defined for the right tail by convention, as the most common extremes of practical interest are maxima, not minima.  Analysis of the earliest arrival concerns the left tail, however, so we will simply multiply the data (in units of days since March 20) by $-1$ and add a constant large enough to make all of the negated values positive, and proceed with the conventional extreme value terminology of referring to annual maxima, rather than minima.  If we imagine choosing a single spatial location and consider the (negative) arrival date $Z$ of each individual bird that passes through, then that variable $Z$ has some cumulative distribution function (\emph{cdf}) $G(z) = P(Z \leq z)$.  Then, making the (unrealistic) working assumption that each individual's arrival is an independent and identically distributed copy of $Z$, the \emph{cdf} of the maximum (negative) arrival from a population of size $n$ is $H(z) = P(\max_{Z_1, \ldots, Z_n} \leq z) = G^n(z)$.

But rather than a single location, we are interested in the random vector $\bZ$, the maximum (negative) arrival at all locations of interest, whose joint \emph{cdf} is $H(\bz) = P(\max_{\bZ_1, \ldots, \bZ_n} \leq \bz) = G^n(\bz)$.  Our goal is to model this distribution of spatial maxima, $H(\bz)$.  The population size $n$ is always going to be unknown, but more fundamentally, our model should remain valid even if the $n$ changes from year to year. A coherent model for $H(z)$, then, would require that $H^{1/n}(\bz)$ be a valid joint distribution for any $n$.  This is exactly the max-infinite divisibility property.  Thus, we require any spatial model for (negative) first arrivals to be max-id.  Similar reasoning applies to random vectors that are not independent across individuals, like arrival dates of migratory birds.

%We use the hierarchical max-id spatial model from \cite{bopp2020} to analyze the Magnolia Warbler first arrival data.  We chose this model over competing models for spatial extremes \citep{reich2012, Huser2018, PADOAN20131} because it can be implemented on large spatial data sets, and because it has flexible spatial tail dependence characteristics, while maintaining the required max-id property.

To define the spatial process $Z(\bs)$ of (negative) first arrivals, now written explicitly as a function of spatial location $\bs$, we first define the process $Y(\bs)$ as a combination of basis functions, 
\begin{equation}
	\label{eqn:Y(s)}
	Y(\bs)=\left \{ \sum_{l=1}^{L}A_l K_l(\bs)^{1/\alpha} \right \}^\alpha.
\end{equation}
Here, $\alpha$ is a parameter that controls the smoothness of the process, and $A_1, \ldots, A_L$ are independent and identically distributed (iid) scaling coefficients.  The scaling coefficients have an exponentially-tilted positive stable distribution, with parameters $\alpha$ and $\theta$ that together control the strength of the spatial tail dependence of the resultant process \citep{bopp2020}.

\begin{figure}
	\begin{center}
		%\vspace*{-4cm}
		\includegraphics[width=1.0\linewidth,clip=true,trim={0 15 0 20	}]{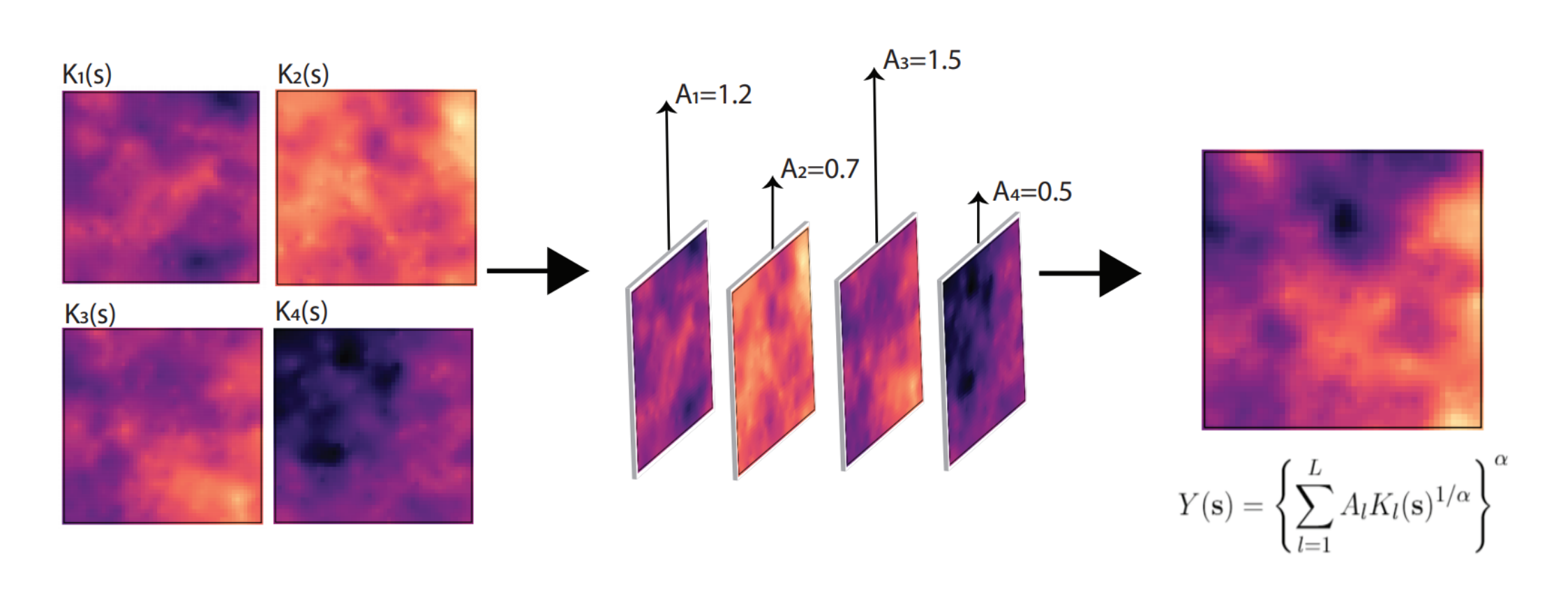} 
		\captionof{figure}{Illustration of the construction of $Y(s)$ using $K_s(s)$ and $A_l$}\label{fig:cartoon}
	\end{center}
\end{figure}

The functions $K_1(\bs), \ldots K_L(\bs)$ in \eqref{eqn:Y(s)} form a collection of $L$ spatial basis functions, which in combination form the shape of the spatial process of first arrivals (see Figure \ref{fig:cartoon}).  We do not know \emph{a priori} what shape of basis functions will result in the best-fitting combination, so we estimate the shape of these functions by assigning them prior distributions based on Gaussian processes.  The construction requires that basis functions be positive and satisfy a sum-to-one constraint, so we specify the priors by transforming independent mean-zero stationary Gaussian processes $\tilde{K}_1(\bs), \ldots, \tilde{K}_{L-1}(\bs)$ as $K_l(\bs)=\text{exp}\left \{ \tilde{K}_l(\bs)/\sum_{i=1}^{L}\text{exp}\left \{ \tilde{K}_i(\bs) \right \} \right \}$, $l=1,\ldots,L,$
%\[
%K_l(\bs)=\text{exp}\left \{ \tilde{K}_l(\bs)/\sum_{i=1}^{L}\text{exp}\left \{ \tilde{K}_i(\bs) \right \} \right \}, \quad l=1,\ldots,L,
%\]
with $\tilde{K}_L(\bs) \equiv 0$ to complete the specification.  In this way, the shape of the basis functions can be learned from the data.

The form of the basis combination \eqref{eqn:Y(s)} resembles a spatial factor model, where traditionally $\alpha$ is taken to be 1 and $A_1, \ldots, A_L$ are iid Gaussian. However, the particular $L^p$ norm (with $p=1/\alpha$) construction of \eqref{eqn:Y(s)} and the particular tilted stable distribution of $A_1, \ldots, A_L$ are the keys to obtaining the desired max-id property \citep{bopp2020}.

To complete the model, we introduce  an everywhere multiplicative ``nugget'' effect $\epsilon(\bs)$ with iid Fr\'echet($1/\alpha$) marginal distributions.  The final spatial max-id model for first arrivals is thus
\begin{equation}
	\label{eqn:Z-with-nugget}
	Z(\bs)=\epsilon(\mathbf{s})Y(\mathbf{s}).
\end{equation}
The nugget effect $\epsilon(\bs)$ represents small-scale variation.  In our case, small-scall variation is particularly relevant because it can capture local habitat or resource variation that is not present in the covariates.  For example, patches of intact forest land or preferred food sources, or even a neighborhood cat, are too small-scale to be captured in county-level data, but can be captured in the model by $\epsilon(\bs)$.

Univariate extreme value theory says that marginally (i.e. at any location), the distribution of the first arrival will converge to a Generalized Extreme Value (GEV) distribution as the population grows.  Therefore, the \cite{bopp2020} model includes a marginal transformation to GEV, from the distribution implicitly defined by \eqref{eqn:Z-with-nugget}, inside the model hierarchy.  This allows flexible modeling of the marginal surfaces, including dependence on covariates, as well as uncertainty propagation between the marginal and joint components of the model.

The GEV($\mu, \sigma, \xi$) is a three-parameter distribution with location parameter $\mu \in \mathbb{R}$, scale parameter $\sigma > 0$, and shape parameter $\xi\in \mathbb{R}$.
%The GEV($\mu, \sigma, \xi$) is a three-parameter distribution with %\emph{cdf}  
%
%\[
%G(z)=\left \{\begin{matrix}
%\text{exp}[-\text{exp}\left \{ -(z-\mu)/\sigma \right \}], & \xi=0, \\ 
%\text{exp}[-\left \{ 1+\xi(z-\mu)/\sigma \right \}^{-1/\xi}_{+}], & \xi \neq 0,
%\end{matrix} \right.
%\]
%where $a_{+}=\text{max}(0,a)$ for location $\mu \in \mathbb{R}$, scale $\sigma >0$, and shape $\xi \in \mathbb{R}$, with $\left \{ z \in \mathbb{R}: (1+\xi(z-\mu)/\sigma) >0 \right \}$,  when $\xi \neq 0$, and $ \left \{ z \in \mathbb{R} \right \}$ when $\xi=0$.
We define $\tilde{Z}(\mathbf{s})=\text{GEV}^{-1}[G_{\mathbf{s}}\left \{ Z(\mathbf{s}) \right \};\mu(\mathbf{s}),\sigma(\mathbf{s}),\xi(\mathbf{s}) ]$ as the (negative) first arrival date on the original observation scale, where $GEV^{-1}\left \{  \cdot ;\mu(\mathbf{s}),\sigma(\mathbf{s}),\xi(\mathbf{s}) \right \}$ is the quantile function of a GEV distribution with parameters $\mu(\mathbf{s}),\sigma(\mathbf{s})$ and $\xi(\mathbf{s})$. $G_{\mathbf{s}}(z)$ is the marginal distribution function of $Z(\mathbf{s})$, which is implicitly defined by the construction \eqref{eqn:Y(s)} and \eqref{eqn:Z-with-nugget}.  Critically, the max-id property of the model is preserved when using this transformation.  Therefore, $\tilde{Z}(\bs)$ is the observed (negative) first arrival date, with GEV($\mu(\mathbf{s}),\sigma(\mathbf{s}))$ marginal distribution, whereas $Z(\bs)$ the same quantity, under the transformation to the marginal distribution induced by \eqref{eqn:Y(s)} and \eqref{eqn:Z-with-nugget}. 

We assume the first arrival process $\tilde{Z}(\mathbf{s})$ is independent across years, given a collection of covariates that may vary in time. Let $\tilde{Z}_t(\mathbf{s})$ to be the process observed at location $\mathbf{s}$ at time $t$. Marginal GEV parameters vary with time via the climate covariates. The spatially varying basis functions are common across time, whereas the scaling coefficients of the basis functions, $A_{l,t}, l=1, \ldots, L$ and $t=1, \ldots, T$, vary with time.  The data did not exhibit evidence of temporal non-stationarity in the spatial dependence parameters $\alpha$ or $\theta$.

\subsection{Model Fitting}

We transform our first arrival dates, of which the extreme arrivals are minima (earliest day of the year) to be maxima by multiplying by $-1$ and adding a large enough constant such that all values are positive. We do all steps of the analysis using the negated data. When plotting predictions, we transform them back to calendar dates for ease of interpretation.

We follow common practice by fixing $\xi(\mathbf{s})$ to be constant in space, with prior  $\xi \sim N(0,100)$, as it is notoriously difficult to discern spatial variation in this parameter \citep{Cooley2007,bopp2020}. We assume a Gaussian process prior for $\mu(\mathbf{s})$, thereby allowing the marginal location parameter of first arrival date to vary flexibly across space. We use spatial and climate covariates to model the mean function of the Gaussian process. We selected the set of covariates with the lowest AIC and BIC scores.  We use goodness of fit measures to select the number of basis functions $L$, with $6, 8, 10, 12$ and $14$ as candidate values for $L$.

\section{Results}

\subsection{Model Comparison}

We used out of sample predictive log scores to select the number of basis function $L$ and to decide between modeling $\sigma(\mathbf{s})$ as a Gaussian process or as a fixed linear model.  For this model selection, we used a subset of 114 counties where at least 10 years of data are present. We then randomly sampled 12 counties as the out of sample set. We fit the model on the rest of the data and evaluated the log likelihood, given the MCMC samples of the model parameters, of the out of sample data set, yielding a log score for each set of posterior samples. We used a 95\% trimmed mean to calculate the average log score for each candidate model. The results are given in Table \ref{table:log-scores} in Appendix \ref{appendix:model-comparison}. The best model under this approach is the model with 8 basis functions and $\sigma(\mathbf{s})$ as a fixed linear model.

\subsection{Final Fitted Model}

We then fitted this model to (negative) first arrival dates from all 869 counties and all years. We used draws from posterior predictive distributions to predict first arrival times at the counties without first arrival observations.

\begin{figure}
	\begin{center}
		%\vspace*{-4cm}
		\includegraphics[width=0.75\linewidth,clip=true,trim={0 20 0 10}]{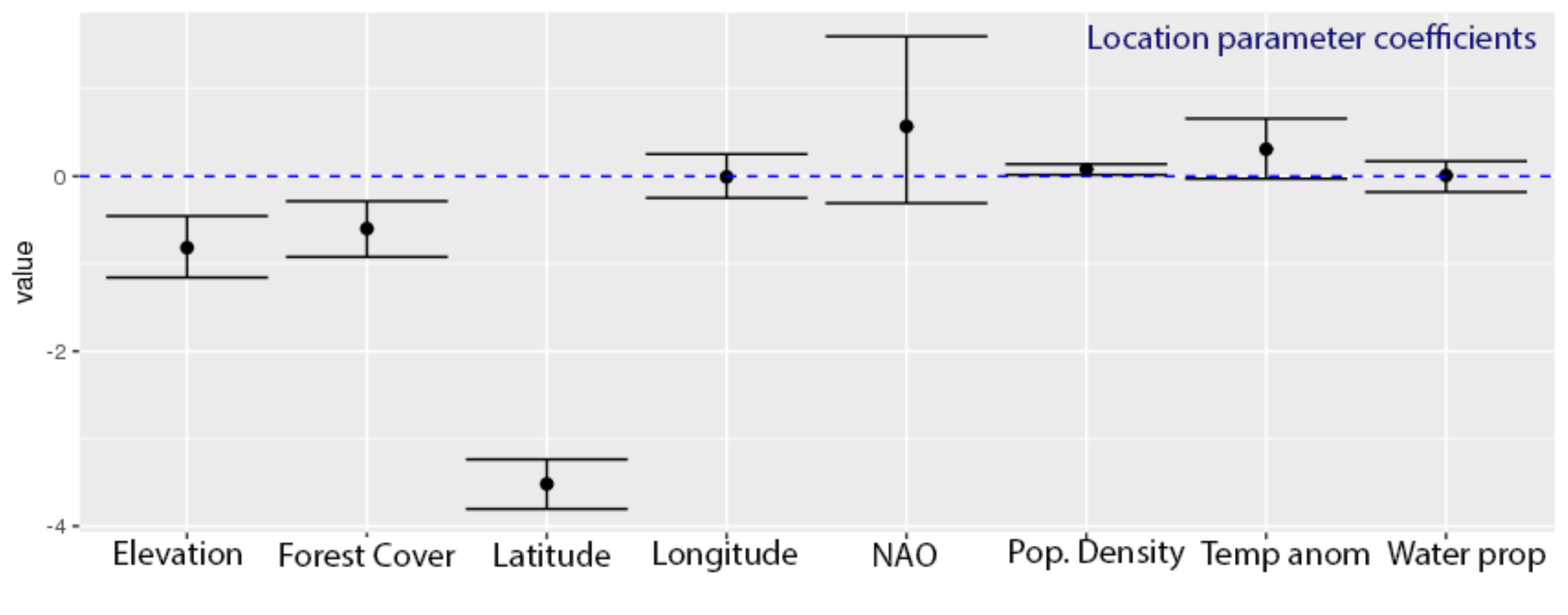} 
		\captionof{figure}{Coefficient estimates and 95\% credible intervals for covariates in the mean function of the location parameter. Transformed (i.e. negated) first arrival data is used here. Larger values for the location parameter correspond to earlier first arrival.}\label{fig:meanlocation}
	\end{center}
\end{figure}

\begin{figure}
	\begin{center}
		%\vspace*{-4cm}
		\includegraphics[width=0.74\linewidth,clip=true,trim={0 20 0 10}]{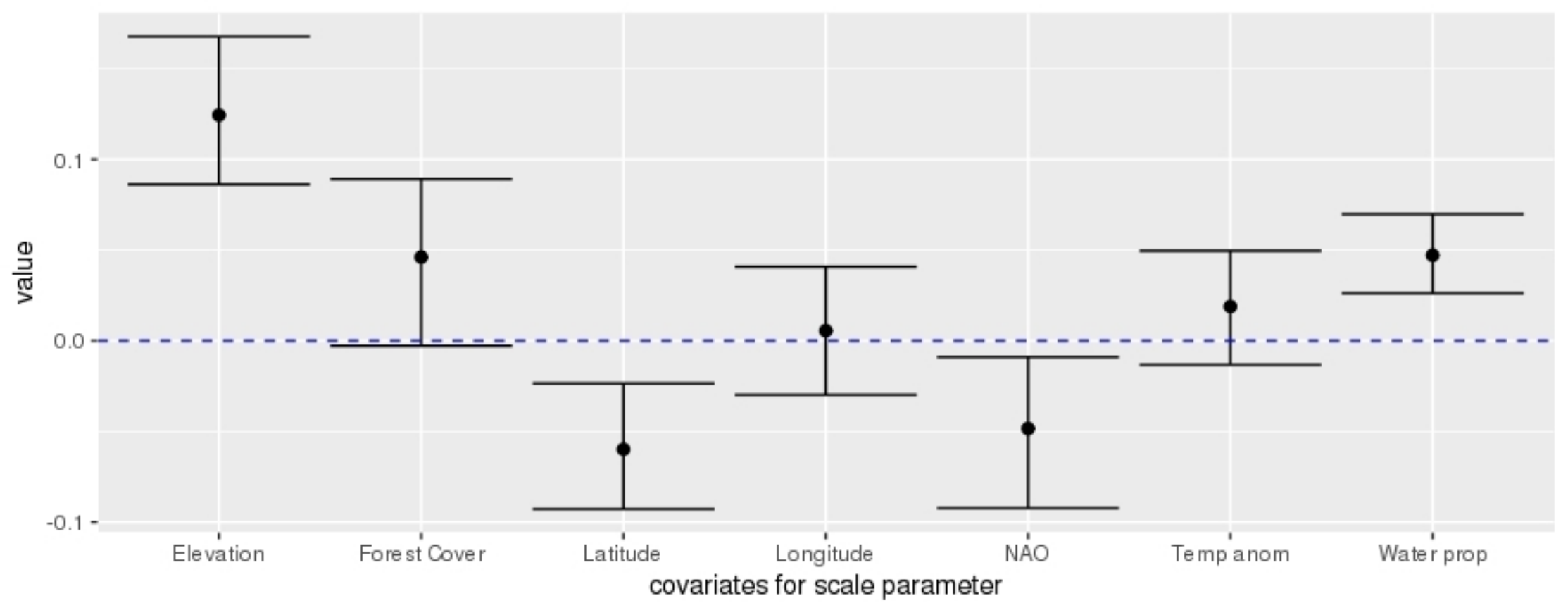} 
		\captionof{figure}{Coefficient estimates and 95\% credible intervals for covariates in the mean function for the scale parameter}\label{fig:meanscale}
	\end{center}
\end{figure}

Posterior means and 95\% credible intervals for $\alpha$, $\theta$ and $\xi$ are 0.3340 (0.3057, 0.3604), 0.00018 (0.000028, 0.00057) and -0.4095 (-0.4435, -0.3784). Posterior means and 95\% credible intervals for the location parameter and scale parameter coefficients are given in Figures \ref{fig:meanlocation} and \ref{fig:meanscale}, respectively. These results correspond to negated first arrival data (i.e. maxima), so that larger values for the location parameter correspond to earlier first arrivals. Latitude has the most negative coefficient from the location parameter covariates, indicating that counties at lower latitudes have earlier arrivals, as expected. Elevation also has a negative coefficient, indicating that regions at higher elevation have later arrivals on average. Forest cover has a negative coefficient. Population density has a slightly positive coefficient. This effect could be a proxy for observation effort, wherein more densely populated regions have more people available to observe the earliest warbler arrivals. Temperature anomaly and NAO both have positive point estimates. Higher values for NAO and temperature anomaly corresponds to warmer weather, which in turns leads to earlier arrival. This too is consistent with what we expect. 

Interpreting the coefficients for the scale parameter covariates is not as straightforward. We can only say that covariates with higher coefficients lead to more variability in first arrivals. For example, regions at higher elevation or more forest cover have higher variability in the first arrival dates.

\begin{figure}
	\begin{center}
		%\vspace*{-4cm}
		\includegraphics[width=0.75\linewidth,clip=true,trim={0 0 0 20}]{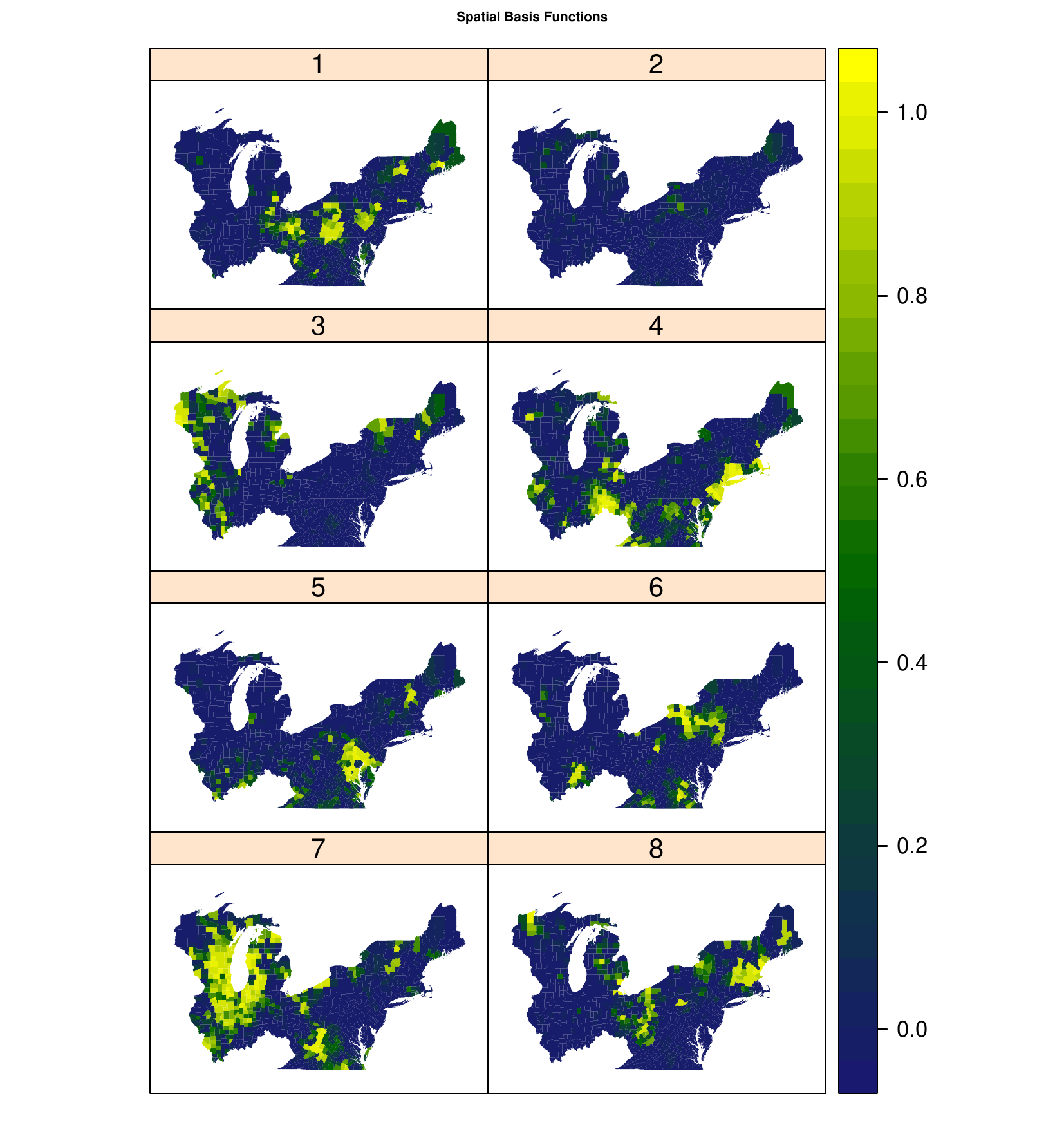} 
		\captionof{figure}{Posterior estimates of the spatial basis functions, ordered by variance of the corresponding random basis coefficients, from largest to smallest.}\label{fig:rbf}
	\end{center}
\end{figure}

Posterior means of the eight random basis functions are plotted in Figure \ref{fig:rbf}. They are ordered by the variance of their corresponding coefficients, in a manner analogous to the ordering of principle components (also known as empirical orthogonal functions).  The first six basis functions account for over 85\% of the variability. These spatial patterns can be interpreted as locations where early or late arrivals tend occur together in the same year, perhaps because they represent migration corridors.

\begin{figure}
	\begin{center}
		%\vspace*{-4cm}
		\includegraphics[width=0.49\linewidth,clip=true,trim={0 220 0 220}]{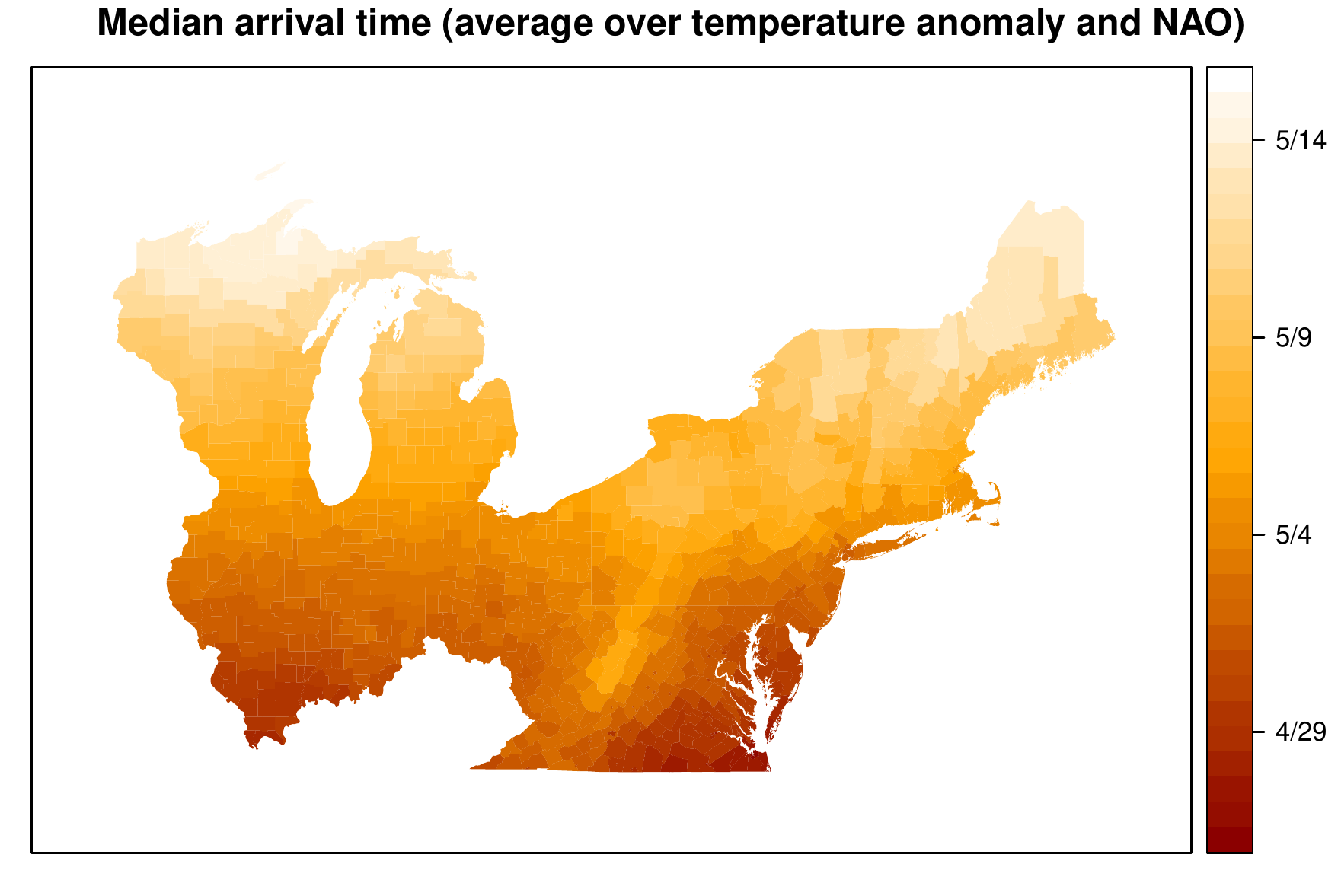} 
		\captionof{figure}{Estimated median first arrival of Magnolia Warblers for 2016--2019, averaging over temperature anomalies and NAO, i.e. predicting using 0 as covariate values for both temperature anomaly and NAO.}\label{fig:median}
	\end{center}
\end{figure}

We also calculated the median first arrival date using posterior sample values for $\mu(\bs)$, $\sigma(\bs)$ and $\xi$, using the formula $z_{0.5}(\bs)=\hat{\mu}(\bs)+\frac{\hat{\sigma}(\bs)}{\hat{\xi}}\left ( \left ( -\log(0.5) \right )^{-\hat{\xi}}-1 \right )$. Here, we averaged over temperature anomaly and NAO and then averaged over the posterior samples and have plotted median first arrival in Figure \ref{fig:median}.  Dark red corresponds to earlier arrival, while light yellow corresponds to later arrival. As evident in Figure \ref{fig:median}, the earliest median first arrival of Magnolia Warblers occurred in Illinois and Virginia. Higher elevation regions of West Virginia and Pennsylvania had later arrivals. The states in the extreme northeast like Maine, New Hampshire, and Vermont, as well as the northern regions of Wisconsin and Michigan, had latest median first arrival.

\subsection{Predictive Maps}

\begin{figure}
	\begin{center}
		%\vspace*{-4cm}
		\includegraphics[width=1.0\linewidth,clip=true,trim={0 10 0 5}]{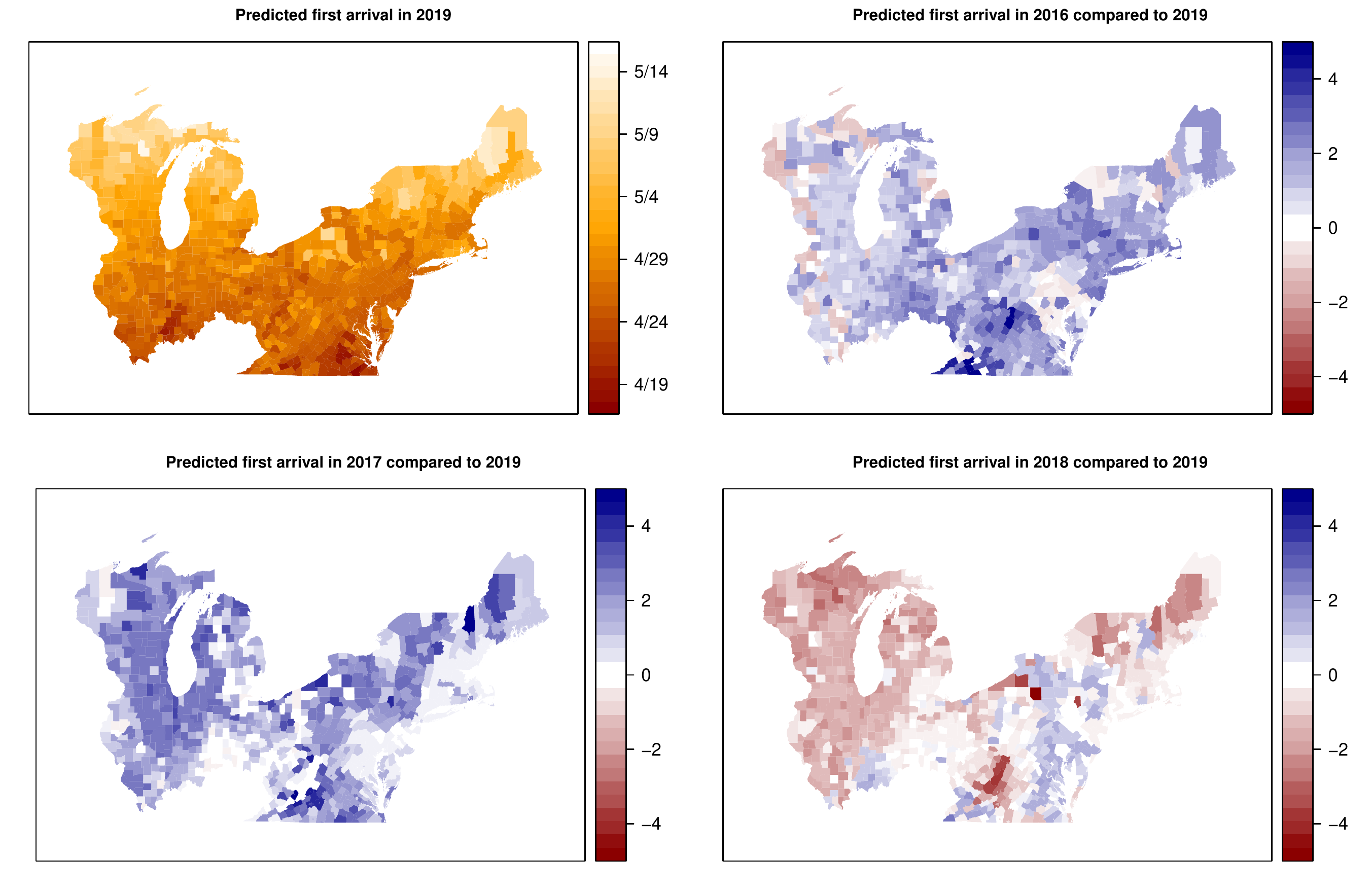} 
		\captionof{figure}{Posterior mean predictive plot for first arrival in years 2016--2019.}\label{fig:postmean}
	\end{center}
\end{figure}

Figure \ref{fig:postmean} gives the mean posterior predictive first arrival for 2019, as well as the difference between 2019 and 2016, 2017, and 2018. In the difference plots, the blue color denotes later arrival while the red color denotes earlier arrival. In 2018 a majority of the counties had earlier arrival compared to 2019 where as in 2017 most counties had later arrival compared to 2019. In 2016 there is a mix of red and blue, with counties in the western region having earlier arrivals while Virginia and parts of the Eastern seaboard having later arrivals. Compared to the median maps, these posterior predictive maps show more small-scale spatial variability, as we are predicting the actual date of first arrival at a given year, rather than an average date. However we see the expected trend of first arrival times being later at higher latitudes compared to lower latitudes. See SI Appendix \ref{sec:MV} Figure \ref{fig:stddev2019} for standard deviations of the posterior predictive first arrival times for 2019.

%\subsubsection{Predictive Maps using Climate model output}

We obtained monthly climate model output from the CMIP5 data portal (\url{https://esgf-node.llnl.gov/search/cmip5/}). The spatial resolution of the dataset is $0.5 \times 0.5$ degrees. We use the climate model GISS-E2-H from NASA's Goddard Institute for Space Studies and picked Representative Concentration Pathway (RCP) scenario `high' (RCP 8.5). RCP8.5 corresponds to high greenhouse gas emissions and is the upper bound of the available RCPs. It is commonly used as a baseline scenario that does not account for any specific climate mitigation strategies \citep{riahi_rcp_2011}. We used ensemble member \texttt{r1i1p1} in this study and extracted data for the time period 2151--2200. 

\begin{figure}
	\begin{center}
		%\vspace*{-4cm}
		\includegraphics[width=1.0\linewidth,clip=true,trim={0 60 0 50}]{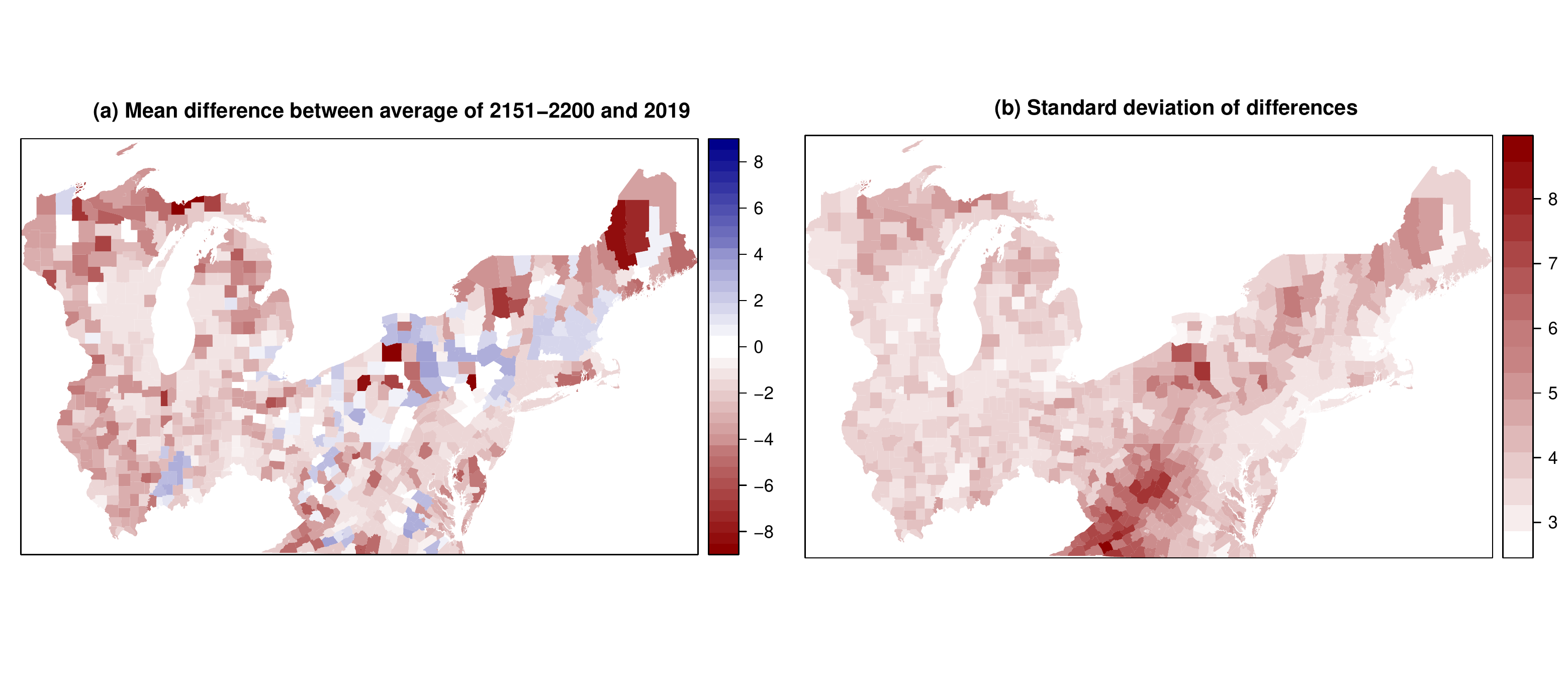} 
		%\vspace*{-7cm}
		\captionof{figure}{(a) Difference in predicted first arrival using base year 2019 for CMIP data averaged over years 2151--2200. (b) Standard deviations of the predictions}\label{fig:meanandsdclimate}
	\end{center}
\end{figure}

Figure \ref{fig:meanandsdclimate} (a) gives the difference between the average projected first arrival dates for 2151--2200 and the first arrival date for 2019 (the last year in our study with actual data). A blue hue denotes later arrival compared to 2019 while, a red hue denotes earlier arrival compared to 2019. As evident in Figure \ref{fig:meanandsdclimate} (a), over 80\% of counties in our region of interest are projected to have earlier arrivals compared to 2019. Some counties in the Appalachian region as well as some with higher forest cover are projected to have slightly later arrivals, as indicated by the light blue hue. Figure \ref{fig:meanandsdclimate} (b) gives the standard deviations of the posterior predictive samples. Higher elevation and higher forest cover regions West Virginia showed high standard deviation values. For difference in first arrival for individual years and their standard deviations, see Figure \ref{fig:diffclimatepred} and Figure \ref{fig:diffclimatesd} respectively in Appendix \ref{sec:MV}.  These predictive maps show a general trend of earlier Magnolia Warbler arrivals under the projected future climate, relative to 2019.

\section{Discussion}

In this study we frame the problem of modeling first arrival of migratory birds as a spatial extremes problem, recognising that first arrival is the realization of the tail of the probability distribution of arrivals. We adapt the hierarchical max-infinitely divisible process model of \cite{bopp2020} to model first arrival of the long distance migrant the Magnolia Warbler, obtaining maps of first spring arrival for the period 2004--2019. We use the posterior predictive distribution from the model to interpolate first arrival dates to counties with missing data in a probabilistically coherent way.

Furthermore, we demonstrated how this method can be used in conjunction with climate model output to produce predictive maps of spring first arrival for future years under projected future climate. Based on the CMIP5 RCP8.5 climate model output we used, we found that first arrival of Magnolia Warblers will be earlier in over 80\% of the counties in the region of interest. We also provided uncertainty estimates for these predictions.

We found that latitude, elevation, and forest cover had negative coefficients when modeling the location parameter of the marginal GEV, indicating later arrival for counties with higher values for these covariates. This is an expected result for latitude but unexpected for forest cover. The forest cover covariate did not distinguish between different types of forests, which could be useful to determine the relationship between forest cover and first arrival more precisely. Both forest cover and elevation had positive coefficients for the scale parameter of the GEV, indicating greater variability for larger values of these variables. 

As expected, temperature anomaly had a positive coefficient for the location parameter, indicating that warmer than average March temperatures result in earlier arrivals of the spring migrants in April. Although the credible interval for NAO for the location parameter overlaps zero, it can be interpreted as having largely positive values, which signals that higher NAO values lead to earlier arrivals. 

This study provides a rigorous basis for studying the first arrival of migratory birds across space and time. It enables prediction of first arrival at unobserved locations while also providing powerful tools to understand the ecosystem response to different climate scenarios.

\section*{Acknowledgements} 
We acknowledge input by Dr. Viviana Ruiz Gutierrez and Dr. Daniel Fink of the Cornell Lab of Ornithology.  The authors gratefully acknowledge support from the US National Science Fountation under  NSF DMS-2001433 and NSF DMS-2015273. Computations for this research were performed on the Pennsylvania State University’s Institute for CyberScience Advanced CyberInfrastructure (ICS-ACI). This content is solely the responsibility of the authors and does not necessarily represent the views of the Institute for CyberScience. This work has no connection to Microsoft Corporation and was conducted by D.A.W. during her doctoral studies, prior to joining Microsoft Corporation.

%%%%%%%%%%%%%%%%%%%%%%%%%%%%%%%%%%%%%%%%%%%%%%%%%%%%%%%%%%%%%%
%%%%%%%%%%%%%%%%%%%%%%%%%%%%%%%%%%%%%%%%%%%%%%%%%%%%%%%%%%%%%%

\appendix

\section{Data Processing} 
\label{appendix:data-processing-all}

\subsection{EBird Data}
\label{appendix:ebird-data}
\begin{figure}
	\begin{center}
		%\vspace*{-4cm}
		\includegraphics[width=0.9\linewidth]{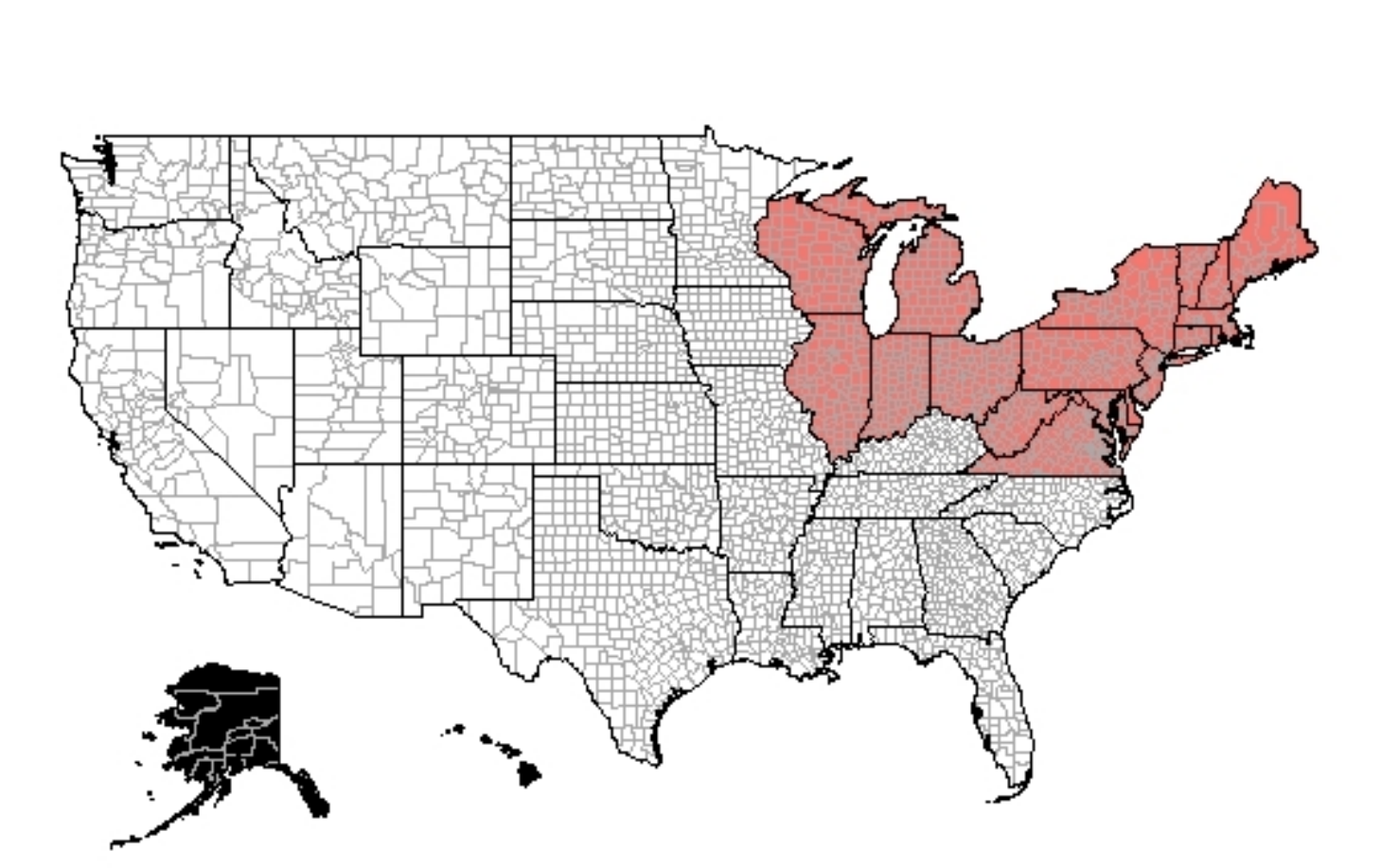} 
		\captionof{figure}{Region of interest in given in red: northeast United States.}\label{fig:reg_interest}
	\end{center}
\end{figure}

We used the citizen data website eBird (\href{www.ebird.org}{www.ebird.org}) to download raw data \citep{Ebirddata} for sightings of the Magnolia Warbler in continental United States. We selected the Magnolia Warbler as an example species because it is a long distance migrant and has a relatively high volume of sightings. The Magnolia Warbler winters in the neotropics. First spring migrants arrive in southern United States in early April. The species is uncommon west of the Mississippi river despite regular vagrancy on the West Coast \citep{BW}. We focused our study on the northeast United States due to the prevalence of Magnolia Warbler sightings in this region. We conducted the study at the county level. The region of interest in given in Figure \ref{fig:reg_interest}. 

\begin{figure}
	\begin{center}
		\centering
		\begin{minipage}{0.50\textwidth}
			\centering
			%\caption{Yearly movements over Elevation}
			% first figure itself
			\includegraphics[width=1.0\textwidth]{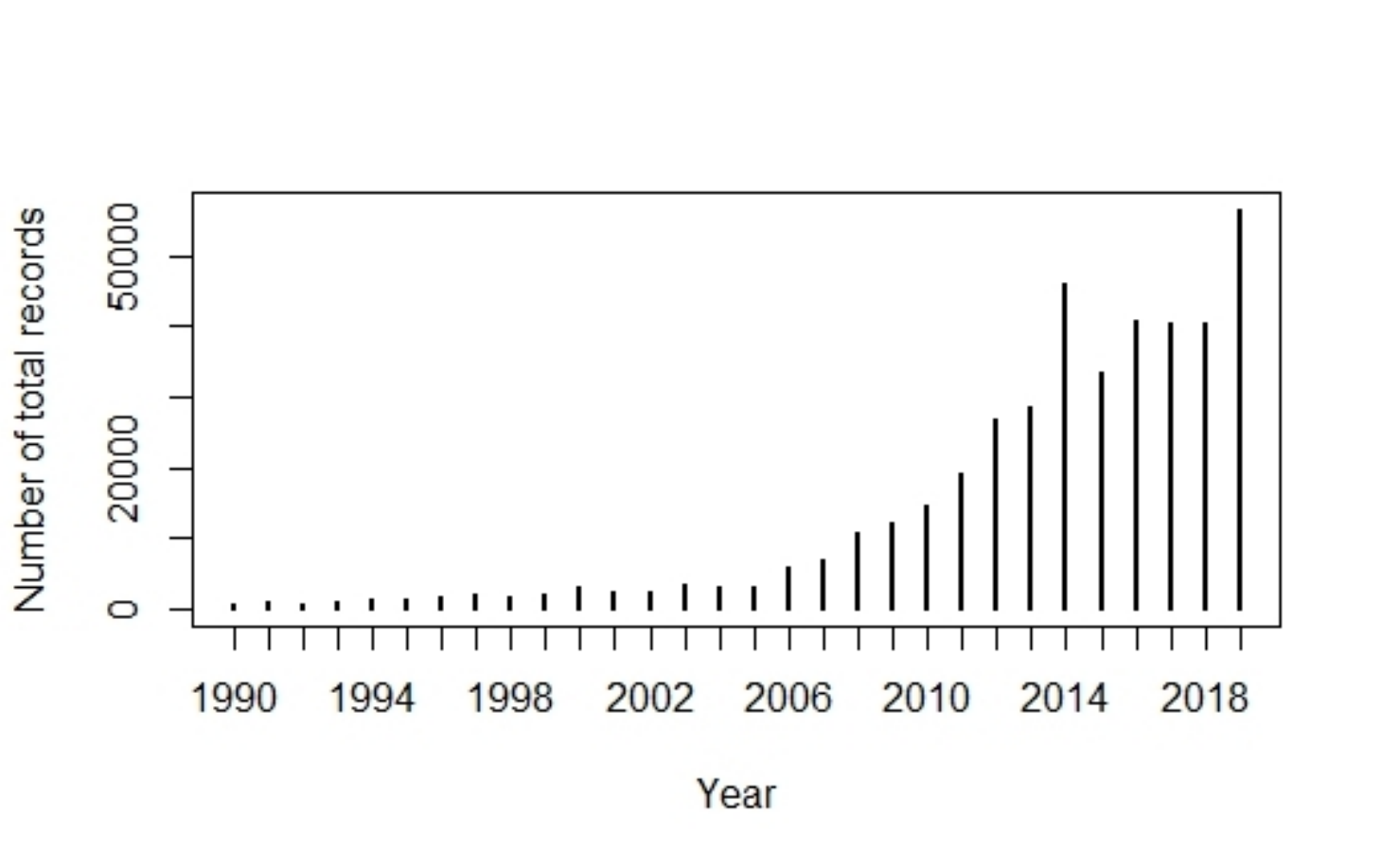} % second figure itself
		\end{minipage}\hfill
		\begin{minipage}{0.50\textwidth}
			\centering
			
			%\caption{Histogram of yearly movements}
			%\vspace*{1.2cm}
			
			\includegraphics[width=1.0\textwidth]{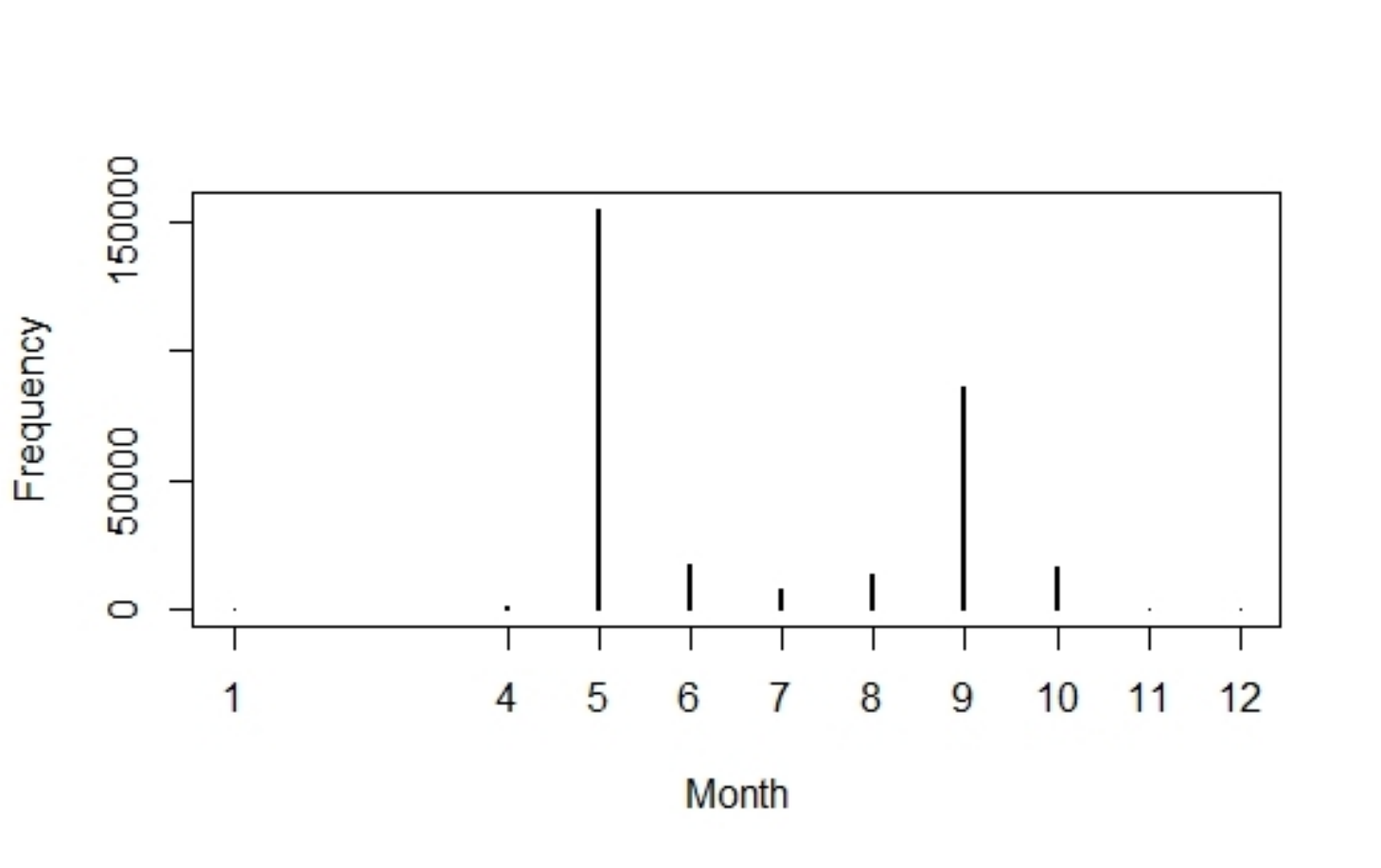}
		\end{minipage}
		%\caption{Visualisation of annual elk movements}
		\captionof{figure}{(a) Total number of records each year. There is a dramatic increase in observation effort in recent years. (b) Earliest spring arrival is in April. Highest number of sightings is in May.}\label{fig:numrecords}
	\end{center}
\end{figure}

The observation effort (i.e. number of active bird watchers) over the years has changed dramatically. Figure \ref{fig:numrecords} (a) gives the number of total records for each year. We only considered data from 2004--2019 in this study, as including data from before this period would lead to many counties having no observations. Figure \ref{fig:numrecords} (b) gives the distribution of data over varying months. There is one observation in January that we discard as carryover from fall migration. Earliest detection of the species in the region of interest is in April, with May having the highest prevalence over the year.  

\subsubsection{First Arrival Data}
\label{appendix:firstarrivaldata}
\begin{figure}
	\begin{center}
		%\vspace*{-4cm}
		\includegraphics[width=0.9\linewidth]{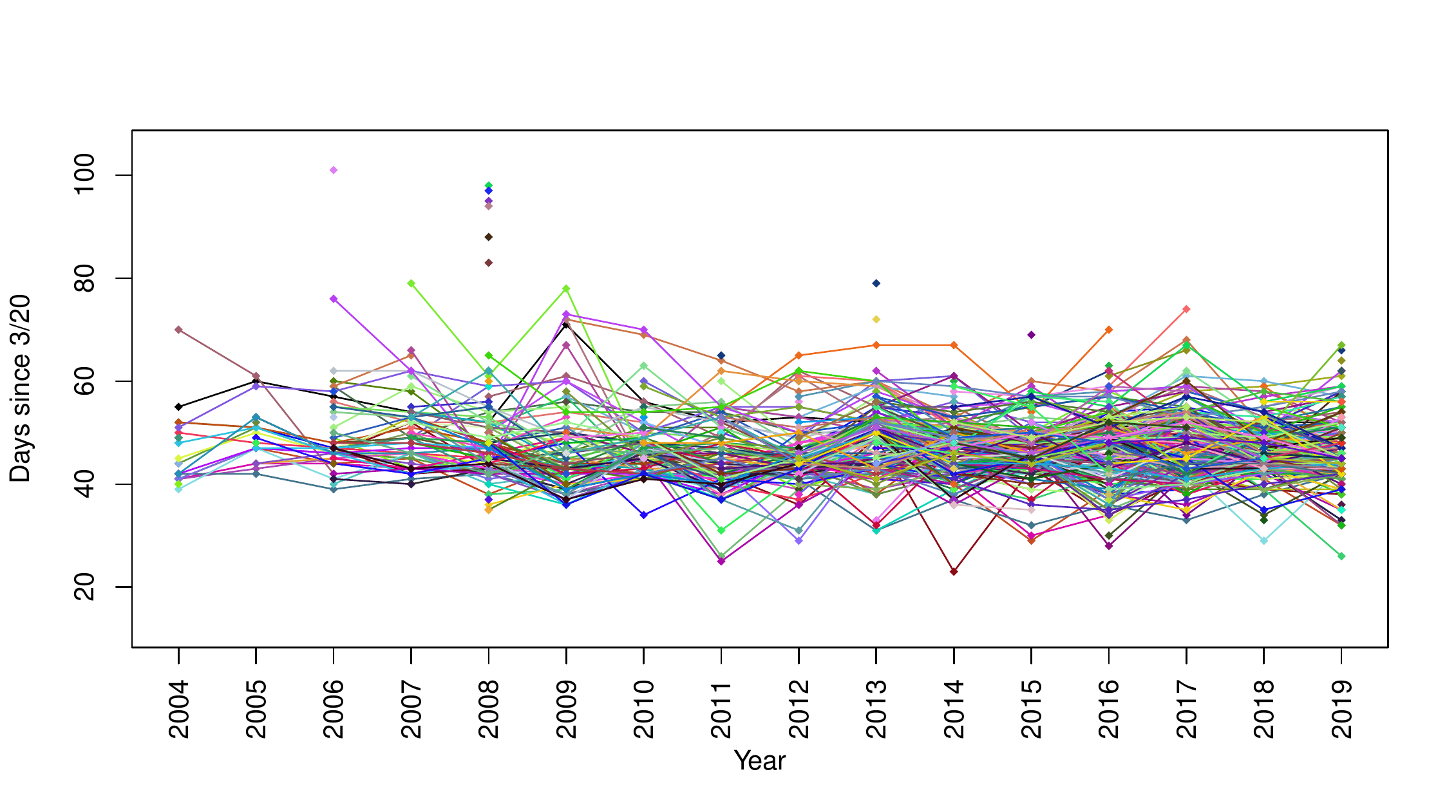} 
		\captionof{figure}{First arrival of Magnolia Warblers colored by county for 2004--2019. First arrival time is only calculated for years in which counties meet the criteria given in section \ref{appendix:firstarrivaldata} First arrival time is calculated as the number of days from March 20. Earliest arrival is 23 days (April 12). Latest arrival is 101 days (June 29). }\label{fig:timeplot}
	\end{center}
\end{figure}

First arrival data can be affected by observation effort and be very noisy \citep{linden}. Our approach in calculating first arrival time is similar to that of \cite{zaifman2017}, who used the same data source. In order to reduce noise we only calculate first arrival time for counties that had at least 12 observations of the bird during the time period March 20 (roughly Spring equinox) and July 20 in a given year. We calculate the first arrival to be the number of days from March 20. \cite{zaifman2017} deleted observations that fell outside of 2.5 standard deviations from the mean for each county. We do not do this because this could delete important outliers. Based on this criteria we identified 402 counties with at least one observation in the 16-year period and 114 counties with at least 10 years of data. Figure \ref{fig:timeplot} gives a visualization of first arrival dates through time, colored by county. There are 869 counties in total in the region of interest. There is a large proportion (80\%) of missing data in the overall dataset.  

We used the Kwiatkowski–Phillips–Schmidt–Shin (KPSS) test \citep{KWIATKOWSKI1992159} to test for temporal non-stationarity for each county. The null hypothesis that temporal non-stationarity exists was rejected for 90\% of the counties. Thus, we did not include temporal non-stationarity in the modeling process.

\subsubsection{Spatial and Climatic Variables}

We use several spatial and climatic variables in our modeling of first arrival dates. For each county, we computed the longitude and latitude by calculating the centroid using the coordinates function of the \texttt{R} package \texttt{sp} \citep{sp_R}. We extracted county-specific elevation values from Terrain Tiles on Amazon Web Services (\url{https://registry.opendata.aws/terrain-tiles/}), which is accessible in \texttt{R} using the function \texttt{get\_elev\_raster} in the package \texttt{elevatr} \citep{elevatr_R}. We obtained elevation raster data from the USDA Forest Service \citep{forestcov} and extracted average county specific values. We also obtained county specific population density and proportion of land covered with water from the 2010 U.S. Census data \citep{census2010}.

\begin{figure}
	\begin{center}
		%\vspace*{-4cm}
		\includegraphics[width=1.0\linewidth]{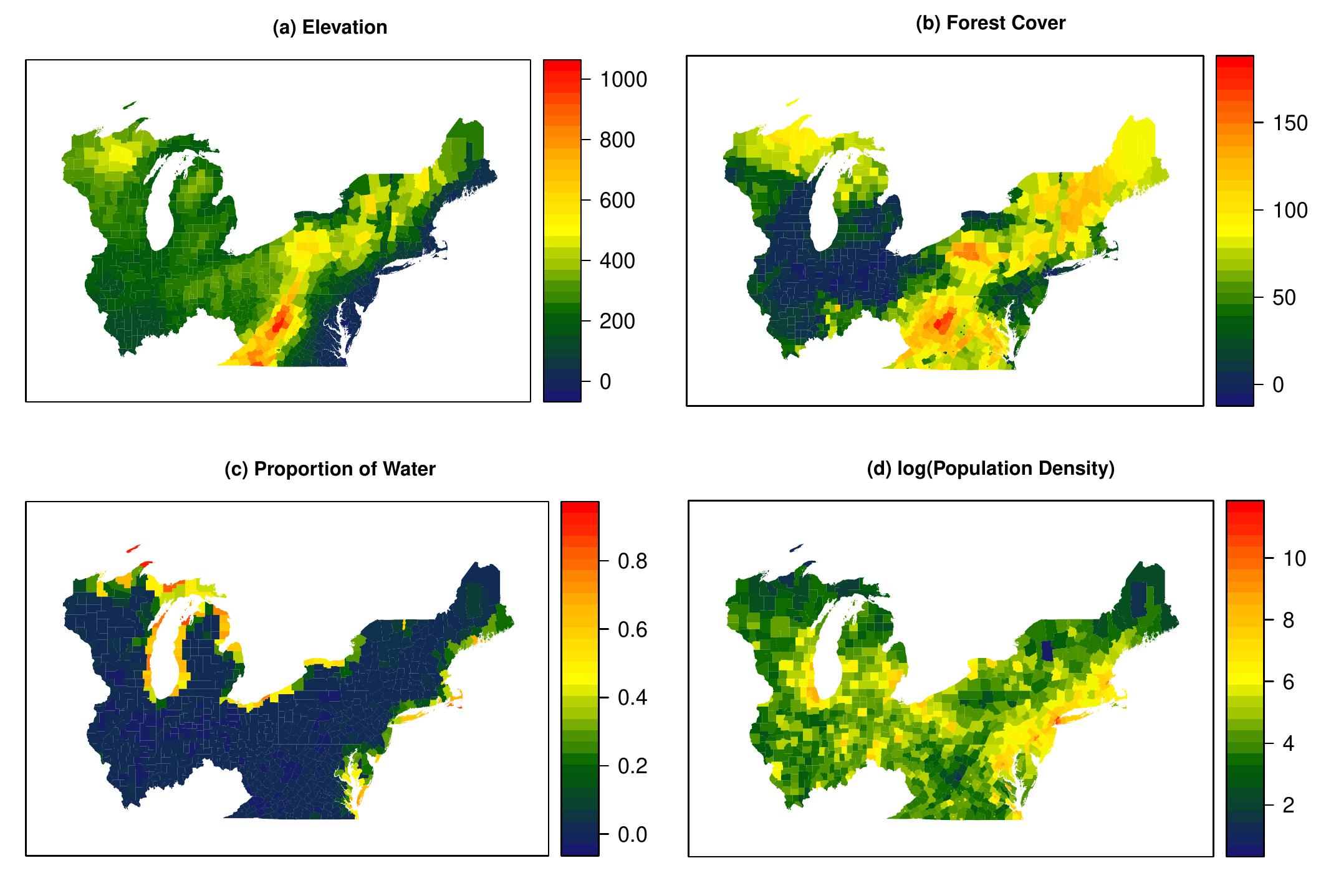} 
		\captionof{figure}{Spatial covariates: Elevation, Forest Cover, Proportion of water and Population density (plotted on log scale).}
	\end{center}
\end{figure}

In addition to these covariates, we also used surface temperature anomaly data on a $5 \times 5$ degree grid for the month of March from the NOAA Global Surface Temperature data set \citep{tempanom}. We extracted county-specific values from this data set. Finally, we obtained station-based North Atlantic Oscillation (NOA) data from \cite{NAO} for the month of March of each year. The station-based NAO is calculated using the difference of normalized sea level pressures between Ponta Delgado (Azores) and Stykkisholmur/Reykjavik, Iceland. Positive values of the NAO index are usually associated with stronger-than-average westerlies over the middle latitudes, warmer winters/cooler summers in Europe and above normal winter temperatures in eastern North America. A negative index is associated with decreased westerlies over the Atlantic and cold dry winters in European areas \citep{NAO, climate_pred_center}. In previous studies done on spring migration of birds in the United States and other European regions, higher values of NAO have been associated with earlier spring arrivals, although this association is weaker for long distance migrants compared to short distance migrants \citep{wilson, Gunnarsson2011, palm2009}. 

\begin{figure}
	\begin{center}
		\vspace*{-8cm}
		\includegraphics[width=0.8\linewidth]{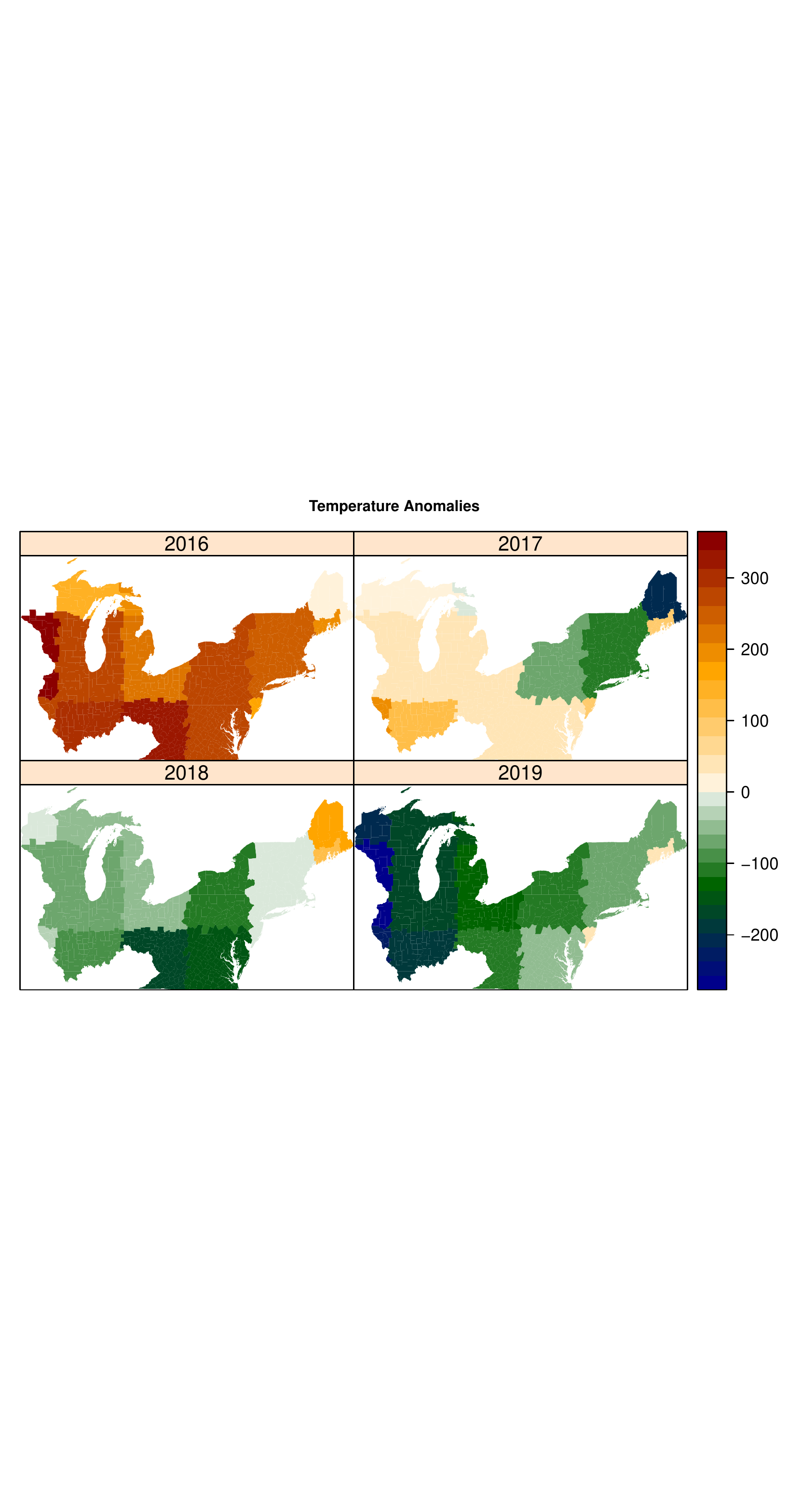} 
		\vspace*{-7cm}
		\caption{Temperature Anomalies for 2016--2019. Given in hundredth of a degree Celcius.}
	\end{center}
\end{figure}

\begin{figure}
	\begin{center}
		%\vspace*{-4cm}
		\includegraphics[width=0.5\linewidth]{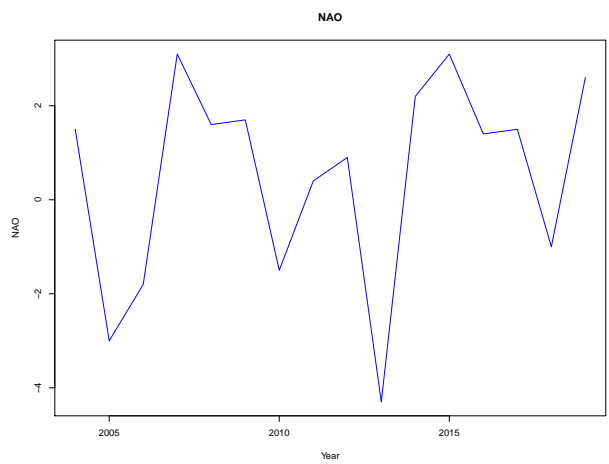} 
		\captionof{figure}{North Atlantic Oscillation (NAO) index for 2004--2019}
	\end{center}
\end{figure}

\section{Re-parameterized Hougaard Distribution}

Following \cite{bopp2020} let $\left \{ A_l \right \}^L_{l=1}$ be a set of i.i.d. random variables distributed as follows

$$A_1, \cdots ,A_L \overset{\text{iid}}{\sim} H(\alpha, \delta, \theta) \; \; \alpha \in (0,1), \delta >0, \theta \geq 0,$$. 

\noindent Where $H(\alpha, \delta, \theta)$ is lighter-tailed, exponentially titled, positive-stable distribution \citep{Hougaard1986}. It has the Laplace transformation

$$\text{E}\left \{ \text{exp}(-sX) \right \}=\text{exp}\left [ -\frac{\delta}{\alpha}\left \{ (\theta +s)^{\alpha}-\theta^{\alpha} \right \} \right ], \;\;\; X \sim H(\alpha,\delta,\theta).$$

Let $f_{PS(x)}$ be the density of $\text{PS}(\alpha)$, a positive stable distribution with Laplace transformation $\text{E}\left \{\text{exp}(-sA)\right \}=\text{exp}(-s^{\alpha}), \;\;\; A \sim PS(\alpha)$ . Then \cite{bopp2020} write the density of $H(\alpha,\delta,\theta)$, $f_{H}$, in terms of $f_{PS}$ as

$$f_H(x)=\frac{f_{PS}\left \{ x(\alpha / \delta)^{1/ \alpha} \right \}(\alpha / \delta)^{1/ \alpha} exp(-\theta x)}{exp(\delta \theta^\alpha /\alpha)}, \;\;\; x>0$$

When $\theta>0$, $f_{PS}$ has the bivariate integral representation

\[f_{PS}(x)=\int_{0}^{1}\frac{\alpha}{1-\alpha}x^{-1/(1-\alpha)}a(\pi u)\text{exp}\left \{ -x^{\alpha/(1-\alpha)}a(\pi u) \right \}\text{d}u\]

where

$$a(v)=\left \{ \frac{\text{sin}(\alpha v)}{\text{sin}(v)} \right \}^{1/(1-\alpha)}\frac{\text{sin}\left \{ (1-\alpha)v \right \}}{\text{sin}(\alpha v)}$$

The parameter $\alpha$ controls the tail decay. Smaller values of $\alpha$ denote heavier-tailed distributions. When $\theta >0$, the gamma distribution with shape $\delta$ and rate $\theta$ is obtained as $\alpha \longrightarrow 0$. 

We use the reparameterization given in \cite{bopp2020}. $\alpha^*=\alpha, \delta^*=(\delta / \alpha)^{1/\alpha}$ and $\theta^*=(\delta/\alpha)^{1/\alpha}\theta$. $\delta^*$ is a scale parameter that does not affect the dependence structure, therefore we set $\delta=\alpha$ and use $H(\alpha,\alpha,\theta)$ (we denote this as $H^*(\alpha,\theta)$ in the main body) without any loss of flexibility. When $\delta=\alpha$ and $\theta>0$, the density $f_H$ is the exponentially tilted form of $f_{PS}$, and $\theta$ exponentially tilting parameter that tapers the tail of $f_{PS}$ at a rate $\theta$.

\section{Details of model fitting}

For both the log-Gaussian process basis functions and the GEV marginal parameter Gaussian processes, \cite{bopp2020} uses a clustering technique to reduce the computational burden, since a Cholesky decomposition of the covariance matrix is required at each iteration. A $k$-nearest-neighbor clustering algorithm is used initially to partition the observation locations into disjoint clusters, and these clusters are fixed for the rest of the algorithm. Block random walk updates are done for each cluster sequentially. For details on the algorithms for the clustering technique, see the Supplemental section of \cite{bopp2020}. 

The dependence parameters $\theta$ and $\alpha$ are assigned priors with $\theta \sim N_{+}(0,100)$ and $\alpha \sim \text{Uniform}(0,1)$. The random basis coefficients $A_{l,t}$ are updated using variable-at-a-time random walk Metropolis steps on the log scale. Variable-at-a-time updating is done for parameters $\alpha$, $\theta$, $\beta_{\psi}$, $\delta^2_{\psi}$, $\rho_{\psi}$, $\xi$, $\delta^2_K$ and $\rho_K$. 

For $\sigma(\mathbf{s})$, we use goodness of fit measures to decide whether it should be modeled as an independent Gaussian process in space or as a simpler linear function of spatial and climate covariates. When $\mu(\mathbf{s})$ and log-scale $\gamma(\mathbf{s}) \equiv \log\left \{\sigma(\mathbf{s})\right \}$ are modeled as Gaussian processes, we specify priors on the coefficients of the mean function of each as $\beta_{i,\psi} \sim N(0,100)$, $\psi \in \left \{ \mu, \gamma \right \}$, $i=1,\cdots, p_{\psi}$, where $p_{\psi}$ is the number of covariates in the mean function of each process. We used a stationary exponential covariance function, $C(h)=\delta^2_\psi \text{exp}(-h/\rho_\psi), h \geq 0$  for $\mu(\mathbf{s})$ and $\gamma(\mathbf{s})$ with half normal priors for $\delta^2_\psi \sim N_{+}(0,100)$ and $\rho_\psi \sim N_{+}(0, \text{max}_{i,j}(\left \| \mathbf{s}_i -\mathbf{s}_j \right \|)^2)$. When $\gamma(\mathbf{s})$ is not modeled as a Gaussian process we simply have $\beta_{i,\gamma} \sim N(0,100), i=1,\cdots, p_\gamma$.

%% Considered moveing the two paragraphs below to main text.
We used a modification of the \texttt{R} package \texttt{stablemix}, included as supplemental material in \cite{bopp2020}, for fitting the model.  The \texttt{stablemix} software performs posterior inference using Markov chain Monte Carlo (MCMC) techniques. In order to select the most appropriate set of covariates for both $\mu(\bs)$ and $\gamma(\bs)$, we fit marginal models with combinations of the covariates using maximum likelihood estimation using the \texttt{R} package \texttt{extRemes} \citep{extremes2016}.  We use an exponential covariance function for the priors on the spatial basis functions and set covariance parameters to be $\delta^2_K \sim N_{+}(0,100)$ and $\rho_{K} \sim N_{+}(0,\text{max}_{i,j}(\left \| \mathbf{s}_i-\mathbf{s}_j \right \|)^2)$. 

For the marginal parameter Gaussian process models, including spatial covariates could lead to mixing difficulties due to possible collinearity between the Gaussian process and the spatial covariates. In order to alleviate this, we transform the Gaussian process to constrain it to be orthogonal to the spatial covariates \citep{hanks_rest_reg2015}. We first sample from the unconstrained process $(\boldsymbol{\eta}^* \sim N(\boldsymbol{\mu}, \bSigma))$ and then apply the transformation $\boldsymbol{\eta}=\boldsymbol{\eta}^*-\bSigma\bX(\bX^{'}\bSigma \bX^{-1})\bX^{'}\boldsymbol{\eta}^*$ where $\bX$ are the covariates used for the orthogonalization, and $\bSigma$ is the covariance matrix of the Gaussian process at the observation and prediction locations. This orthogonalization approach is known as ``conditioning by Kriging'' \citep{cressie_1993}. Care must be taken when performing inference on the regression coefficients in the mean function because constraining the spatial random field to be orthogonal to the fixed effects can result in increased Type-1 error rates \citep{hanks_rest_reg2015}. We use the adjustment $\tilde{\bbeta}=\bdelta-(\bX^{'}\bX)^{-1}\bX^{'}\boldsymbol{\tilde{\eta}}, \boldsymbol{\tilde{\eta}} \sim N(\mathbf{0}, \bSigma)$,  where $\tilde{\boldsymbol{\beta}}$  are the adjusted posterior samples, $\boldsymbol{\delta}$ are the un-adjusted posterior samples, $\boldsymbol{\tilde{\eta}}$ is the orthagonalized Gaussian process. The resulting $\tilde{\boldsymbol{\beta}}$ have Type-1 error rates near nominal levels \citep{hanks_rest_reg2015}. Using the orthogonalization method of \cite{hanks_rest_reg2015} in the implementation of the model of \cite{bopp2020} is a novel approach that improves computation in the fitting of spatial extreme models with covariates. 

 We ran the chain for 110,900 iterations, discarding the first 45,000 iterations as burn-in and using a thinning interval of 50.  
 
\section{Model Comparison}
\label{appendix:model-comparison}

The highest log score in Table 1 is -441.62, which corresponds to the model with 8 random basis functions and the scale parameter of the marginal GEV not specified as a Gaussian process.

\begin{table}[h]
	\caption{Model Comparison: 95\% trimmed mean log scores calculated on the held out dataset. }
	\begin{center}
		\begin{tabular}{|l|c|l|}
			\hline
			\multicolumn{1}{|c|}{\multirow{2}{*}{\textbf{Model No.}}} & \multicolumn{2}{c|}{\textbf{Scale Parameter of GEV modeled as a GP}}             \\ \cline{2-3} 
			\multicolumn{1}{|c|}{}                                    & \multicolumn{1}{l|}{\textbf{No. of Random Basis Functions}} & \textbf{Log score} \\ \hline
			\textbf{1} & 6                                                            &  -488.92                  \\ \hline
			\textbf{2} & 8                                                            & -567.20                   \\ \hline
			\textbf{3} & 10                                                           &  -499.31                  \\ \hline
			\textbf{4} & 12                                                           &   -593.82                 \\ \hline
			\textbf{5} & 14                                                           & -924.87                   \\ \hline
			\multirow{2}{*}{\textbf{}}                                & \multicolumn{2}{c|}{\textbf{Scale Parameter of GEV not modeled as a GP}}         \\ \cline{2-3} 
			& \multicolumn{1}{l|}{\textbf{No. of Random Basis Functions}} & \textbf{Log score} \\ \hline
			\textbf{6} & 6                                                            & -528.84                   \\ \hline
			\textbf{7} & 8                                                            &  -441.62                   \\ \hline
			\textbf{8} & 10                                                           &  -500.59                  \\ \hline
			\textbf{9} & 12                                                           & -698.54                   \\ \hline
			\textbf{10} & 14                                                           & -623.82                   \\ \hline
		\end{tabular}
	\end{center}
\label{table:log-scores}
\end{table}

\section{Standard Deviation of posterior predictive first arrival times for 2019}

\begin{figure}
	\begin{center}
		%\vspace*{-4cm}
		\includegraphics[width=0.7\linewidth]{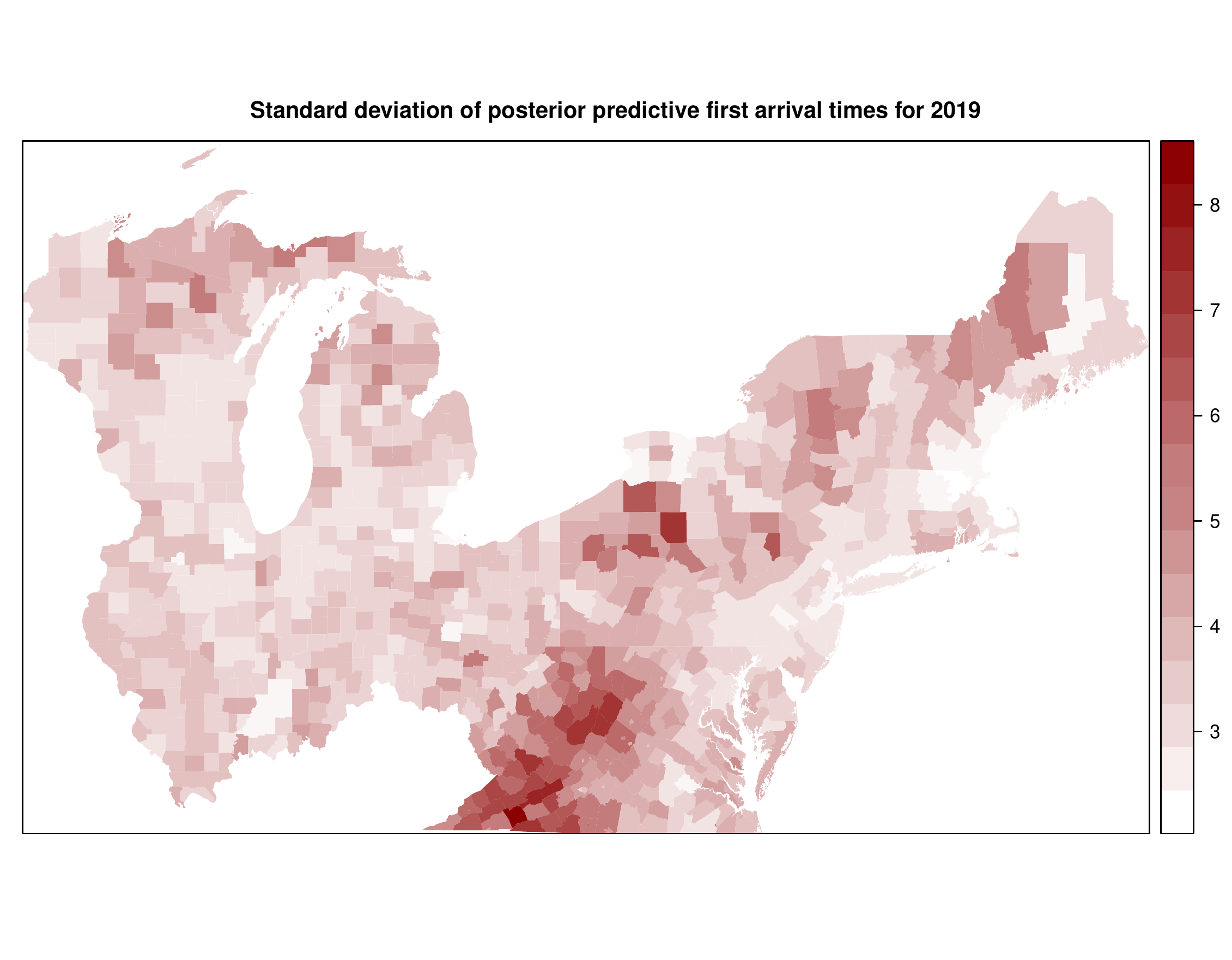} 
		\captionof{figure}{Standard Deviation of posterior predictive first arrival times for 2019.}\label{fig:stddev2019}
	\end{center}
\end{figure}

Figure \ref{fig:stddev2019} gives the standard deviation of the posterior predictive first arrival times for 2019. 

%[hbt!]

\section{Figures of climatic variables and predicted first arrival for Climate model Data}
\label{sec:MV}

\begin{figure}
	\begin{center}
		%\vspace*{-4cm}
		\includegraphics[width=0.8\linewidth]{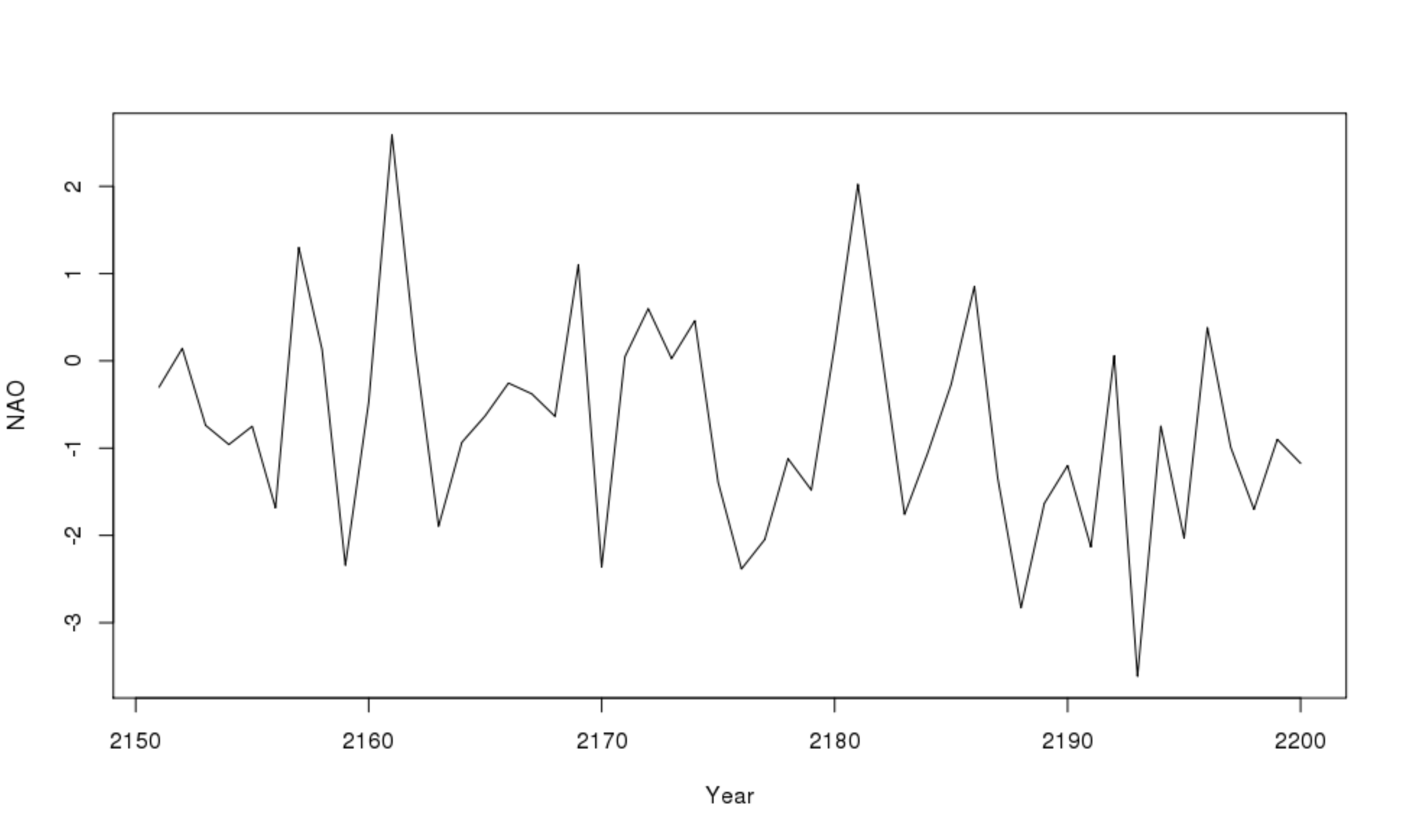} 
		\captionof{figure}{North Atlantic Oscillation (NAO) index using CMIP5 RCP8.5 climate model output for 2151--2200}\label{fig:naoclimate}
	\end{center}
\end{figure}

\begin{figure}
	\begin{center}
		%\vspace*{-4cm}
		\includegraphics[width=0.8\linewidth]{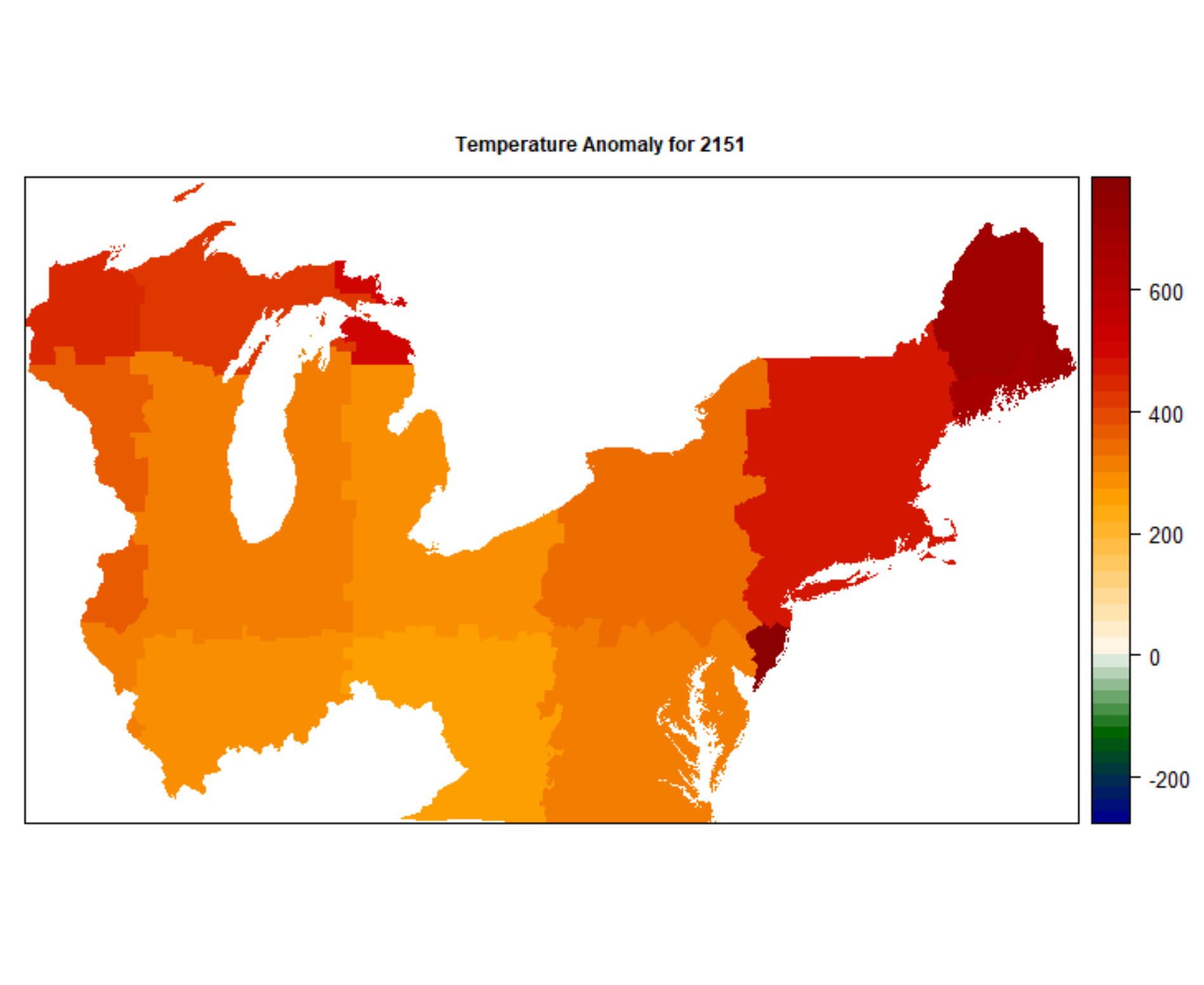} 
		\captionof{figure}{North Atlantic Oscillation (NAO) index using CMIP5 RCP8.5 climate model output for 2151--2200}\label{fig:tempanomclimate}
	\end{center}
\end{figure}

%\BAS{I think this paragraph should go in an appendix.}
We first obtained sea level pressure and surface temperature values for the month of March of each year from the above climate model and calculated NAO and temperature anomaly based on baseline values, following \cite{NAO} and \cite{tempanom}, respectively. Calculated NAO and temperature anomaly for the year 2151 are given in Figures \ref{fig:naoclimate} and Figure \ref{fig:tempanomclimate}, respectively. We used MCMC samples of the parameters $\alpha$ and $\theta$ from the final fitted model to draw samples for $A_{l,t}$ from the unconditional distribution of the exponentially tilted positive stable distribution with parameters $\alpha$ and $\theta$. Then, we obtained samples of the posterior predictive distribution for first arrivals using MCMC samples from the final fitted model of $\alpha$, $\theta$, $\xi$, the 8 basis functions, the Gaussian process for the location parameter, location and scale parameter coefficients, and covariates with updated NAO and temperature anomaly values. We used the same spatial coefficients and updated only the climatic variables using CMIP5 data. NAO values were negative many of the years under consideration while temperature anomaly values were significantly higher than the baseline zero, signaling clear evidence of warming climate.

NAO calculated based on CMIP5 RCP8.5 climate model data for 2151--2200 is given in Figure \ref{fig:naoclimate}. Temperature anomaly calculated for 2151 is given in Figure \ref{fig:tempanomclimate}. 

\begin{figure}
	\begin{center}
		\vspace*{-7cm}
		\includegraphics[width=0.7\linewidth]{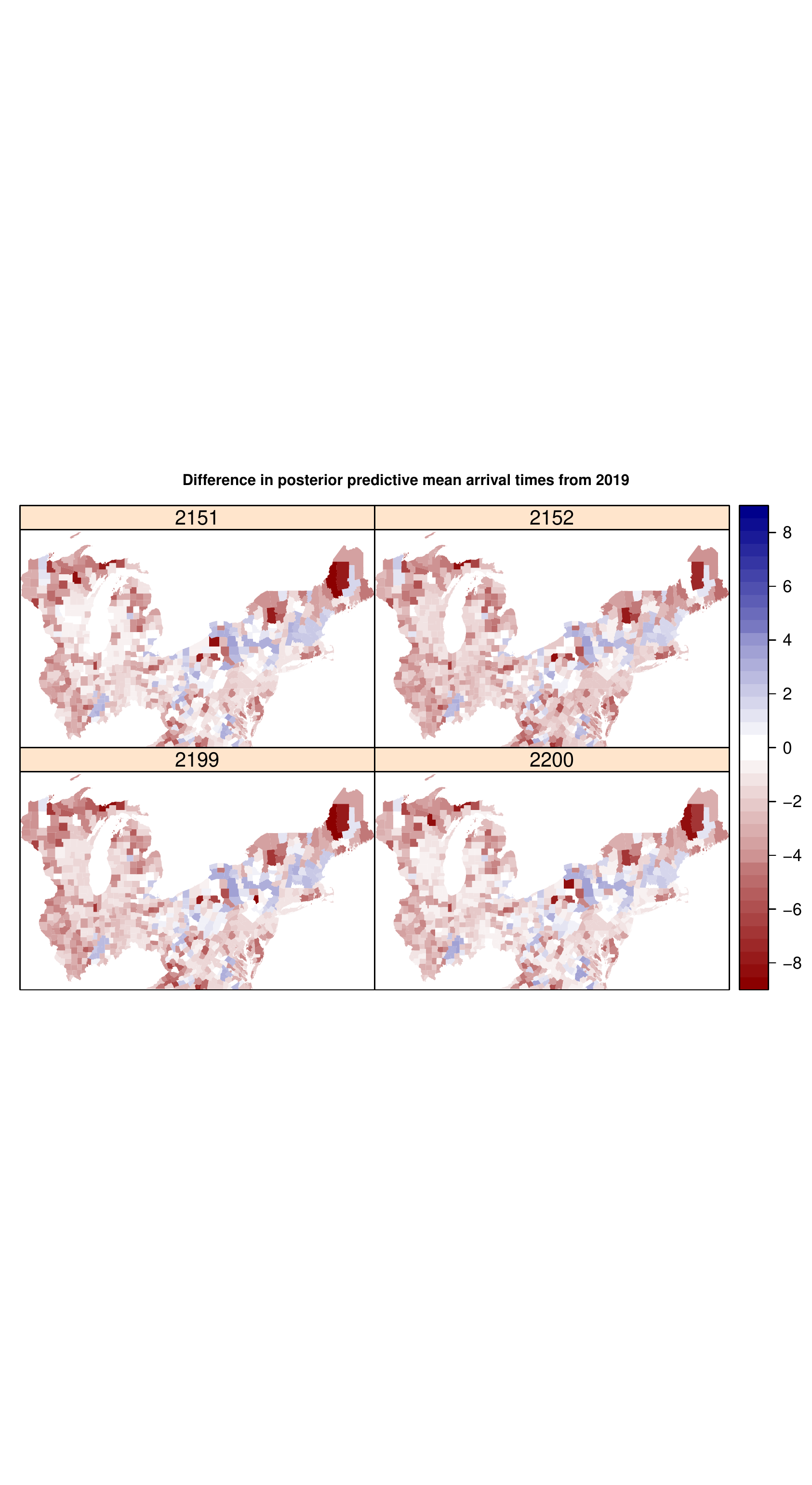} 
		\vspace*{-7cm}
		\captionof{figure}{Difference in predicted first arrival using base year 2019 for CMIP data from years 2151, 2152, 2199, and 2200.}\label{fig:diffclimatepred}
	\end{center}
\end{figure}

\begin{figure}
	\begin{center}
		\vspace*{-5cm}
		\includegraphics[width=0.7\linewidth]{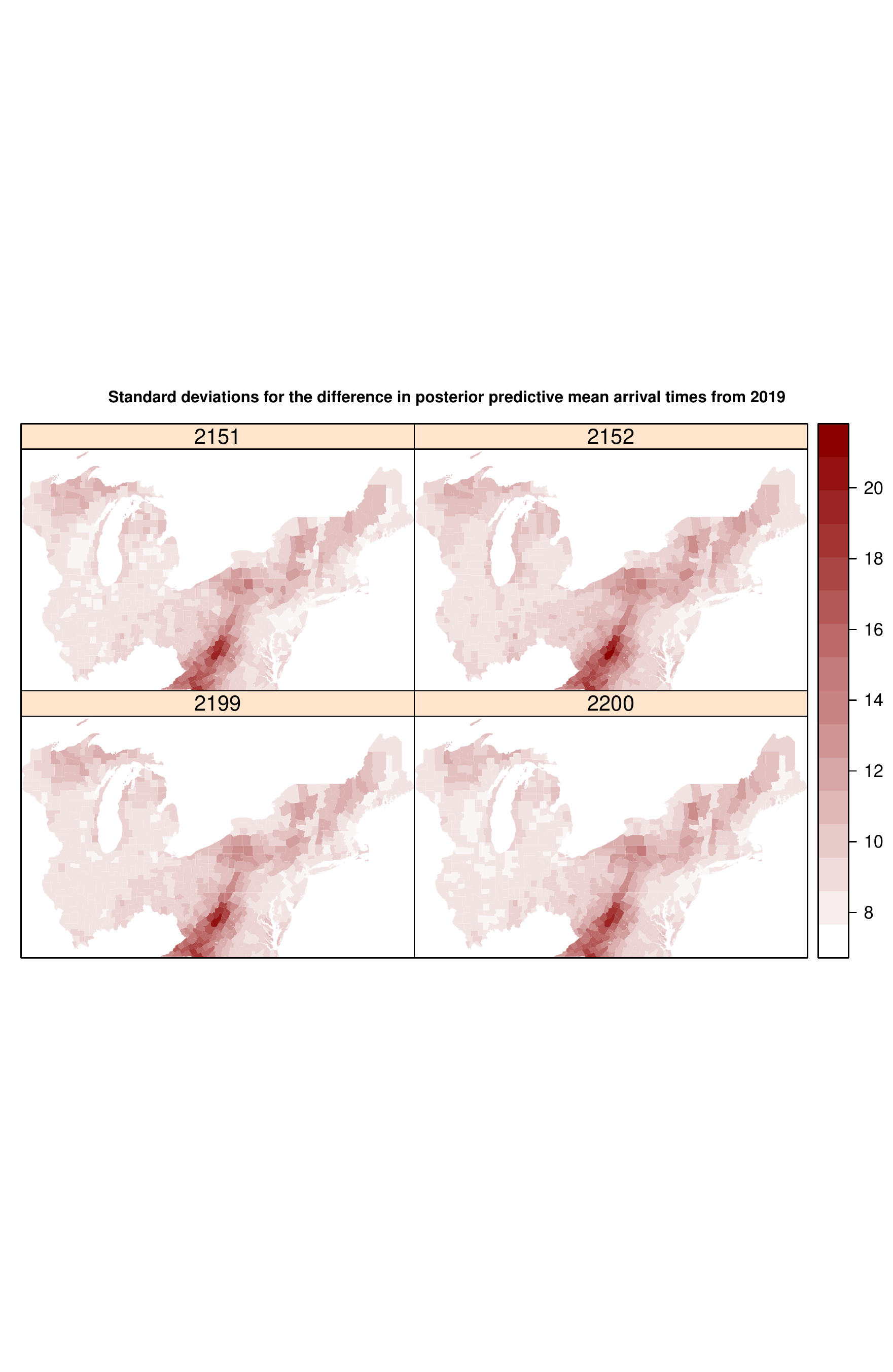} 
		\vspace*{-5cm}
		\captionof{figure}{Standard deviation of the difference in predicted first arrival using base year 2019 for CMIP5 data from years 2151, 2152, 2199 and 2200.}\label{fig:diffclimatesd}
	\end{center}
\end{figure}

Figures \ref{fig:diffclimatepred} gives the mean difference in predicted first arrival from 2019 for years 2151, 2152, 2199, and 2200. Figure \ref{fig:diffclimatesd} gives the standard deviation of these estimates. 

\section{Future Work}

%\BAS{If we want to make cuts, this first paragraph is a good candidate.}
This method can be used to develop a tool that can give predictions of first arrival at counties with no sightings given sightings at a small number of counties. This could be useful to ornithologists who study behaviors of these species and also to bird watching enthusiasts. In conjunction with this it would be useful to look at ways in which the computing burden could be further reduced. The clustering approach used by \cite{bopp2020} is useful in increasing efficiency of the MCMC sampler. We did not run the full model with all 869 counties for variable selection and model selection purposes due to the computing burden. Having a tool that gives conditional predictions at unobserved location would be most helpful if those predictions can be provided in a timely manner.

By framing the analysis of migratory first arrivals as a spatial extremes problem, and applying a hierarchical max-infinitely divisible process model, we were able to obtain spatially complete maps of first arrival dates in past years, as well as project future first arrival dates based on climate model output, all in a probabilistically coherent way.  In this study we only considered one species: the Magnolia Warbler. However this model could be used (individually) to study multiple species and to compare predicted first arrival as well as interpret coefficients to understand which species are adapting better to climatic changes. 

Comparing the adaptability of short distance migrants to long distance migrants is commonly seen in literature studying first arrival of migratory birds \citep{kullberg_change_2015,tottrup_local_2010}. In this study we used a long distance migrant. Extending this study to short distance migrants might need more careful consideration as identifying first arrival is difficult in areas where there is a higher probability of seeing the species year around. Carefully selecting the region of interest is vital in such a study, so as not to select areas where there is a high probability of bird sightings year round. 

In this study we considered a time period of 2004--2019 for 869 counties. We had a large number of missing values in the data. However this is not a problem as the MCMC sampler draws posterior predictive draws for missing values at each iteration. The model will benefit from having more time replicates as the observation effort continues each year. 

We fitted the model at the county level, aggregating over observations made within each county. It was necessary to do aggregation at some level since observations are not made at the same locations every year and we need fixed spatial locations to fit the model proposed by \cite{bopp2020}. We used counties as our observation unit since it is easier to obtain values of covariates aggregated at this level. However, not all counties are similarly sized, which could lead to our heuristics being very stringent on smaller counties. Another approach to do this could be to use an evenly spaced grid over the region of interest.  

NAO values calculated for the climate model data was mostly negative over the time period of interest (2151--2200), although temperature anomaly values showed significant warming of eastern North America. This leads to a larger discussion of the station-based NAO index. In this study we use the station-based NAO index of \citep{NAO} due to the ease in computing NAO values for climate model data using sea level pressure. However, as \citep{NAO} states, this index is based on two individual stations, and their pressure readings can be noisy due to small scale meteorological phenomena unrelated to NAO. NAO based on rotated principal component analysis is a better measure, since it is based on the entire flow field and not only by the difference at two locations \citep{climate_pred_center}. However, it is more difficult to compute the RCPA based NAO using raw sea level pressure values. 

In this study our primary goal is to demonstrate the use of these methods. We do not attempt to collect the most comprehensive list of covariates for modeling the spring arrival of the Magnolia Warbler. EBird data that is accessed through the associated \texttt{R} package \texttt{ebirdst} \citep{ebirdst} contains a large number of potential predictor variables. For example, there are eight different types of forest classification (evergreen needleleaf forests, deciduous broadleaf forests, etc.) included, which could be used to better understand preference of the migratory species. We did not use these covariates in our analysis as they are only available for years 2014--2019. %As the observation effort increases as seen in Figure 2 (a) \BAS{use cross reference}, and capturing potentially associated covariates become easier a large number of these covariates could be used in the modeling process in the future. 

In our study we used a set of heuristics for determining if first arrival should be included at a specific county in a specific year. The goal of these heuristics was to reduce noisy observations, especially in early years when observations effort was lower. These heuristics could be too conservative and exclude sightings of vagrants by experienced observers. Alternatively, it might be advantage to take advantage of the Checklist Calibration Index (CCI) \citep{kelling2015}, which provides a rating for the reliability of each eBird observer.

\bibliographystyle{unsrtnat}
\bibliography{sources}

\end{document}